\documentclass[prd,superscriptaddress,amsfonts,amssymb,amsmath,showpacs,twocolumn]{revtex4-2}
\usepackage{subcaption}
\usepackage{bm}
\usepackage{amsfonts}
\usepackage{braket}
\usepackage{latexsym}
\usepackage[latin1]{inputenc}
\usepackage{graphicx}
\usepackage{amsmath}
\usepackage{mathtools}
\usepackage{palatino}
\usepackage{ragged2e}
\usepackage{mathpazo}
\usepackage{textcomp}
\usepackage{float}
\usepackage{booktabs}
\usepackage{dcolumn}
\usepackage{hyperref}
\usepackage[english]{babel}
\usepackage[autostyle]{csquotes}
\hypersetup{colorlinks,citecolor=red}
\usepackage{amsmath}
\usepackage{xcolor}
\usepackage{orcidlink}
\usepackage{graphicx}
\usepackage{subcaption}
\usepackage{float}
\usepackage{commath}

\def\jnl@style{\it}
\def\aaref@jnl#1{{\jnl@style#1}}

\def\aaref@jnl#1{{\jnl@style#1}}

\def\aj{\aaref@jnl{AJ}}                   
\def\apj{\aaref@jnl{ApJ}}                 
\def\apjl{\aaref@jnl{ApJ}}                
\def\apjs{\aaref@jnl{ApJS}}               
\def\apss{\aaref@jnl{Ap\&SS}}             
\def\aap{\aaref@jnl{A\&A}}                
\def\aapr{\aaref@jnl{A\&A~Rev.}}          
\def\aaps{\aaref@jnl{A\&AS}}              
\def\mnras{\aaref@jnl{Mon.~Not.~Roy.~Astron.~Soc.}}             
\def\prd{\aaref@jnl{Phys.~Rev.~D}}        
\def\prc{\aaref@jnl{Phys.~Rev.~C}}  
\def\prl{\aaref@jnl{Phys.~Rev.~Lett.}}    
\def\qjras{\aaref@jnl{QJRAS}}             
\def\skytel{\aaref@jnl{S\&T}}             %
\def\ssr{\aaref@jnl{Space~Sci.~Rev.}}     
\def\zap{\aaref@jnl{ZAp}}                 
\def\nat{\aaref@jnl{Nature}}              
\def\aplett{\aaref@jnl{Astrophys.~Lett.}} 
\def\apspr{\aaref@jnl{Astrophys.~Spawith the current cosmological paradigm.ce~Phys.~Res.}} 
\def\physrep{\aaref@jnl{Phys.~Rep.}}      
\def\physscr{\aaref@jnl{Phys.~Scr}}       
\def\commat{\aaref@jnl{Comm.~Math.~Phys.}}              
\def\science{\aaref@jnl{Science}}               
\def\cqg{\aaref@jnl{Classical Quant.~Grav.}}            
\def\jpcs{\aaref@jnl{JPCS}}                                     
\def\ijmpd{\aaref@jnl{Int.~J.~Mod.~Phys.~D}}                    
\def\grg{\aaref@jnl{Gen.~Relat.~Gravit.}}               
\def\rpp{\aaref@jnl{Rep.~Prog.~Phys.}}          
\def\npa{\aaref@jnl{Nucl.~Phys.~A}}        
\def\lrr{\aaref@jnl{Living Rev.~Rel.}}                   
\def\jcap{\aaref@jnl{J.~Cosmology Astropart.~Phys.}}    
\def\rmp{\aaref@jnl{Rev.~Mod.~Phys.}}   
\def\epjc{\aaref@jnl{Eur.~Phys.~J.~C}}

\allowdisplaybreaks[1]

\begin{document}

\color{black}

\title{Reconstruction of a dark energy model for the Dirac-Born-Infeld scalar field with the Hubble and DESI data via Gaussian process}

\author{Sayantan Ghosh\orcidlink{0000-0002-3875-0849}}
\email{sayantanghosh.000@gmail.com}
\affiliation{Department of Mathematics, Birla Institute of Technology and Science, Pilani, Hyderabad Campus, Jawahar Nagar, Kapra Mandal, Medchal District, Telangana 500078, India.}
\author{Gaurav N. Gadbail\orcidlink{0000-0003-0684-9702}}
\email{gauravgadbail6@gmail.com }
\affiliation{Faculty of Symbiotic Systems Science, Fukushima University, Fukushima 960-1296, Japan.}
\author{P.K. Sahoo\orcidlink{0000-0003-2130-8832}}
\email{pksahoo@hyderabad.bits-pilani.ac.in}
\affiliation{Department of Mathematics, Birla Institute of Technology and Science, Pilani, Hyderabad Campus, Jawahar Nagar, Kapra Mandal, Medchal District, Telangana 500078, India.}
\author{Kazuharu Bamba \orcidlink{0000-0001-9720-8817}}
\email{bamba@sss.fukushima-u.ac.jp}
\affiliation{Faculty of Symbiotic Systems Science, Fukushima University, Fukushima 960-1296, Japan.}


\begin{abstract}
In this study, we reconstruct the dark energy (DE) as a Dirac-Born-Infeld (DBI) scalar field from the Hubble dataset (32 CC + 26 BAO) and the DESI dataset using the Gaussian process (GP). As the GP is a non-parametric and model-independent way to reconstruct a function and its derivative using the data, our reconstruction of the DE equation of state, the DE density parameter, and the potential does not assume any particular model of cosmology. Using Monte Carlo realizations of the GP-reconstructed expansion history, we derive a posterior estimate of the Hubble constant, obtaining $H_0 = 69.53 \pm 2.68$ km s$^{-1}$ Mpc$^{-1}$. This method offers a fully model-independent estimate of $H_0$, relying only on data and GP priors, and provides an unbiased intermediate value useful for reassessing the Planck-SH0ES tension.
Using the reconstructed profiles of the scalar potential as a function of the field $\phi$, along with their associated uncertainties, we perform a chi-square curve fitting procedure to assess the viability of four different scalar field potentials, such as Exponential, Power-law, Free Field (quadratic), and Higgs-like potential. This allows us to identify which potential best fits the reconstructed data. We also employ MCMC analysis to place quantitative constraints on the model parameters associated with each potential. Furthermore, we do a $\chi^2$ analysis for all four potentials and comment on the goodness of the fit for each of them.  Finally, we discuss possible generalizations of our model-independent framework and outline the phenomenological implications of our findings.

\end{abstract}
\maketitle

\section{Introduction} \label{sec:I}
Since the discovery of Cosmic Microwave Background (CMB) radiation in 1965 \citep{cmb}, it has been more or less confirmed that our universe originated from a very hot, dense state, popularly known as the Hot Big Bang model or just the Big Bang model. It was soon confirmed that CMB radiation is around $3K$, which is remarkably consistent with the decade-old calculation carried out by Gamow and his collaborators \citep{gamow}. Also, it has been shown by them that the Big Bang Nucleosynthesis (BBN) exactly predicts the abundance of heavier materials and also predicts the Hydrogen-to-Helium ratio in the universe with remarkable accuracy. However, there have been some problems with the standard Big Bang model of cosmology, such as the Horizon problems, the Flatness problem, and the monopole problem. It was Alexei Starobinsky \citep{star} and Alan Guth \citep{ALAN} who independently proposed the ``inflation theory" as a way to avoid these problems. Later, in 1983, Andrei Linde \citep{Linde} solved some of the shortcomings of the inflationary model (mainly the initial value problem) and introduced ``chaotic inflation," which makes a broader class of scalar field potentials applicable for the inflationary paradigm. One can think of inflation as a scalar field that acts for a very short time, gives rise to a very rapid expansion (de Sitter-like solution) for a very short time, and decays. Such a rapid expansion in the very early universe could naturally solve all the above-mentioned problems. Even though there are plenty of scalar field potentials and other types of non-canonical scalar fields, such as the K-essence field, the DBI field, and the Generalized DBI field, it has not been possible to pinpoint one single type of scalar field or mechanism that is responsible for the inflation. The best one can do is give a bound on the potential forms via various phenomenological constraints such as the scalar-to-tensor ratio or anisotropies in the CMB.
 
  The second paradigm change in cosmology came with the discovery of late-time acceleration, as noted independently by A. G. Riess \citep{late1} and S. Perlmutter \citep{late2}. By analyzing high-redshift Type Ia supernovae, they showed that the universe is not just expanding, but also accelerating. An obvious explanation for such a situation is that the cosmological constant ($\Lambda$) is responsible for the de Sitter-like expansion. Although the $\Lambda$CDM model is extremely phenomenologically sound and can explain a wide range of phenomenological and observable results (including the latest Planck satellite observations \citep{P18}) based on very few free parameters. However, it has some serious downsides, especially since no explanation is given on how the term $\Lambda$ originates. First and foremost, it was argued by Zel'dovich \citep{ZEL} that maybe $\Lambda$ is just the zero point energy of the vacuum in the quantum fields; however, it has been soon realized that a the energy contribution of the first loop correction using the Planck energy cutoff would give a discrepancy of $\Lambda$ with the observation of the order of $10^{120}$. There is no natural explanation for such a huge discrepancy in contemporary quantum field theory. One natural way to bypass this issue is given by S. Weinberg \citep{WENB} that perhaps the cosmological constant has not been uniform throughout the age of the universe. There is a secondary ``quintessence" scalar field which can naturally explain the late-time acceleration and such a tiny value of observed $\Lambda$.
  
 There is extensive literature available on dark energy, including the review by Copeland et al. \cite{Copeland2006}, which covers most of the contemporary models of dark energy. On the phenomenological side, a review of dark energy in relation to observational constraints, such as those from weak lensing, BAO, and supernovae, has been conducted by Weinberg et al. \cite{Weinberg2013}. The concept of dark energy in the context of modified gravity was reviewed by Bamba et al. \cite{Bamba2012}. For an updated perspective that incorporates more recent observational constraints, see Joyce et al. \cite{Joyce2015}. Finally, the review by Frusciante and Perenon \cite{Frusciante2020} provides a detailed discussion of dark energy arising from various effective field-theory approaches.
 
  In this article, our focus is on the Dirac-Born-Infeld (DBI) scalar field, a special type of $k$-essence field which naturally occurs in open string theory when one takes the low-energy effective Lagrangian of D-Branes \citep{green,polchinski}. In particular, it has been noted that \citep{mazumdar,sen1,sen2,sen3} tachyon condensation of the non-BPS branes leads to such a DBI potential, which can be used in the context of cosmology as the energy scale is very similar to that of pre-inflationary scenarios. The detailed study of the DBI field as inflation has been done by Padmanabhan \citep{paddy} and Gibbons \citep{gibbons1}. Moreover, the DBI field as a candidate for the quintessence field has also been explored by various authors, as it can not just reproduce the late-time acceleration but also give a value that is very consistent with the observation, and last but not least, it has a very sound physical ground. Alternative ways of obtaining the DBI field from other forms of string theory have been reviewed by Gibbons \citep{gibbons2}. The study of the DBI field in a late-time acceleration context was done by Bagla et al. \citep{bhagla}, while Gorini et al.\citep{gorini}
offered an alternative way of visualizing the DBI field as a modified Chaplygin gas. It has also been noted by T. Padmanabhan and T. Roy Choudhury \citep{paddy2} that the DBI field can not only explain dark energy (DE) but also dark matter (DM), making it an ideal candidate for the study of late-time cosmology. It should also be noted that apart from string theory, one can get the DBI field using the generalized Chaplygin gas equation of state as shown by Gorini \citep{gorini}.

The analytical study of the DBI or Tachyon fields is usually done using dynamical system analysis. As the Klein-Gordon equation for DBI fields is highly nonlinear in nature, it is very difficult to study using the exact solutions. Therefore, one typically incorporates the fixed point analysis using the dynamical system method. The first study using dynamical system analysis was conducted by Copeland \citep{copeland1} and Aguirregabiria \citep{aguirregabiria}. However, they only used the inverse square type potential in order to find the tracker solution. Soon after that, it was extended to more general potentials beyond the inverse square potentials by Fang and Lu \citep{fang2010a}. Quiros \citep{Quiros} first tackled even more general functional forms, including the $sinh(\phi)$ type potential. Guo \citep{guoexp} has studied the exponential potentials in detail and given a very solid motivation as to why exponential potentials, apart from the power law potentials, are a very good choice for such cases. We have followed these leads to choose the power law and exponential potentials, and we have included the more general potentials such as free field and Higgs potentials.

It is also noteworthy that, as demonstrated by Silverstein and Tong~\citep{tong}, when a D3-brane is considered moving toward the horizon of AdS space, one obtains a generalized DBI action in the strong coupling regime -- contrasting with earlier analyzes conducted in the weak coupling limit. In this strong coupling scenario, the DBI field receives additional contributions due to the motion of the D3-brane, leading to a modified Lagrangian of the form
$\mathcal{L}_{\text{GDBI}} = -\frac{1}{f(\phi)} \left( \sqrt{1 + f(\phi) \, \partial_\mu \phi \partial^\mu \phi} - 1 \right) - V(\phi)$. This shows that the DBI fields are a good candidate to study cosmology from string theory, as both high- and low coupling ones can get the DBI field. This gives us a wide range of flexibility in choosing the model that is consistent with the data and phenomenologically sound.
We should also mention that the DBI field or Tachyon fields have already been explored in the context of modified gravity, such as minimally coupled \citep{mincouple} and non-minimally coupled \citep{nonmincouple} $f(R)$ gravity, as well as in the context of $f(Q)$ gravity \citep{medbi}.


Here, we have used the Gaussian process (GP) algorithm to construct a non-parametric reconstruction of the scalar field potential profile within the DBI framework, agnostic to the analytical form of $\mathcal{V}(\phi)$. To do this, we used both combined Hubble data set \citep{Jimenez2002,Moresco2016,Gaztanaga2009,Alam2017} and the latest realized DESI data set \citep{desi,desi1} to constrain the DBI field potential using the GP. GP is a non-parametric algorithm that can enable the reconstruction of a function using the data instead of assuming a specific functional form. The detailed motivation for the algorithm comes from the machine learning algorithm, which is discussed in detail in \citep{Rasmussen2005}. For the cosmological context, it is a very powerful tool, as it is well known that one can reconstruct the form of DE in the form of $H(z)$ and its derivatives via Sahni and Starobinsky's work \citep{sahnireconstruction}. Even though constraining the form of DE or quintessence scalar field potential from data is very common via MCMC or other Bayesian-type analysis, the problem lies with the fact that almost all the Bayesian algorithms choose a form of the potential and find the constraint on the free parameters from the data by performing MCMC analysis. The advantage of GP is that, by design, it is nonparametric, so it does not rely on any prior knowledge of cosmographic quantities such as $w$ or $V(\phi)$, etc. The study of cosmography, that is, the time evolution of the cosmic expansion in a model-independent way using GP, was first done by Shafieloo, Kim, and Linder \citep{lindergaussian}. Subsequently, Holsclaw et al.~\citep{Holsclaw2010a,Holsclaw2010b} pioneered the nonparametric reconstruction of the dark energy equation of state $w(z)$ using Gaussian process modeling applied to supernova data. Following this work, the reconstruction of the model-independent DE profile using GP was done by Seikel et al.~\citep{Seikel2012} and was later optimized and expanded to include various priors \citep{Seikel2013}. Mehrabi et al. \citep{Mehrabi2021} have used the GP and more data sets from Supernova to study the cosmography, showing the robustness of the GP. The reconstruction of the model-independent quintessence scalar field potential using GP was performed in \citep{jesusgaussian,niugaussian}. It is also worth noting that, in recent years, considerable efforts have been made to validate the cosmological swampland conjecture using GP. E. Elizalde et al. have shown \citep{elizaldeswampland1,elizaldeswampland2} that one can verify the consistency of the swampland conjecture \citep{vafa} with the scalar field potential reconstructed from the GP. In a similar model-independent spirit, Yang~\citep{Yang2020} employed GP to study coupled dark energy scenarios and their compatibility with swampland criteria. Significant work is also done on reconstruction of DE and scalar field potentials using GP in the context of modified symmetric teleparallel gravity by G. Gadbail et al. \citep{gaurav1,gaurav2}.

In our work, using the GP reconstruction of $H(z)$, we generated Monte Carlo realizations at each redshift to propagate observational uncertainties into the Hubble constant. Flattening these samples yields a posterior distribution for $H_0$, which is then fitted with a Gaussian profile. This procedure ensures that GP-based uncertainties are consistently reflected in the final constraint on $H_0$. Additionally, we examine the DE behavior in a model-independent manner by reconstructing the evolution of $w_{\phi}$ and $\Omega_{\phi}$ using GP.
Furthermore, we have constructed the kinetic term ($\dot{\phi}^2$) and potential ($V(\phi)$) from the cosmographic parameters ($H(z)$ and its derivatives). Then we have numerically solved the ODE and used GP to reconstruct the potential. We have also taken four different types of DBI potentials such as exponential, power-law, free field, and Higgs field, and have done an MCMC analysis to constrain the free parameters of the potential. Finally, we used the $\chi^2$ test to comment on the goodness of the analysis. In general, this work provides a complete and systematic model-independent analysis of the DBI scalar field and constraints on potentials using the GP. It should be noted that the latest released DESI data set has already been used to reconstruct $f(T)$ gravity using genetic algorithms by R. E. Ouardi et al. \citep{genetic}, and to reconstruct the Om-diagnostic using the GP by P. Mukherjee et al. \citep{gaussianom}. However, to the best of our knowledge, this is the first paper that uses DESI data to tackle the DBI field and find the constraints on the coefficients of the DBI potential. It should also be noted that the interest in fixing the parameters of the potentials from the data is not only important for phenomenological purposes but also has a deeper theoretical necessity. It is well known that string theory is one of the most promising theories of quantum gravity, yet there is no consistent de-Sitter vacuum solution in string theory in the context of cosmology. Such an unexpected result has been handled by C. Vafa's Swampland conjectures \citep{vafa}. The main idea of Swampland is that there are certain gravity theories that are consistent with the quantum theory (that the low-energy gravity theory has a full UV-complete theory); these belong to the ``landscape". However, there are some other theories, such as theories based on the de-Sitter vacuum belonging in the ``swampland", in a sense, these theories cannot have a reasonable UV-complete quantum theory. Based on this paradigm, Vafa has given some conjectures, such as the Swampland Distance Conjecture (SDC), the de Sitter Conjecture (dSC) and the refined de Sitter Conjecture (RdSC), which in principle gives a very tight bound on \( \left|\frac{\nabla_{\phi}V}{V}\right| \).  It has been shown by Elizalde and Khurshudyan \citep{elizaldeswampland1} that one can indeed verify the Swampland conjecture, or place bounds on the constants of the potentials, using GPs. Later, they have also extended this verification to modified \( f(R) \) gravity formulations \citep{elizaldeswampland2}.

The structure of this work is as follows: In Section \ref{sec2}, we briefly discuss the physical motivation behind the DBI field and its basic mathematical formalism. In Section \ref{sec4}, we perform a model-independent reconstruction of the Hubble function and its derivatives using the CC+BAO+DESI datasets. Using this reconstruction, we further reconstruct the DE equation of state and the DE density parameter. Section \ref{sec5} focuses on the reconstruction of the scalar field potential through the GP, where we also employ chi-square analysis for model selection and MCMC analysis to constrain the model parameters. Finally, we summarize and conclude our results in Section \ref{sec6}.

\section{Physical motivation and MATHEMATICAL FORMALISM OF THE DBI FIELD THEORY}\label{sec2}
 $\,\,\,\,\,$The DBI action arises in open string theory as an effective low-energy description of the dynamics of the D-brane \citep{green,polchinski}. In particular, tachyon condensation in non-BPS D-branes \citep{mazumdar,sen1,sen2,sen3} generates a DBI-type scalar field with potential applications to cosmology. Padmanabhan \citep{paddy} and Gibbons \citep{gibbons1} demonstrated that DBI fields in Friedmann-Lemaitre-Robertson-Walker (FLRW) backgrounds exhibit inflationary behavior, while subsequent studies explored their role in late-time acceleration \citep{bhagla,gorini,paddy2}. These works suggest DBI fields could provide a string-theoretic origin for cosmic acceleration without ad hoc scalar-field introduction.

Dynamical systems analysis has proven invaluable for studying cosmological scalar field dynamics \citep{copeland1,aguirregabiria}. By recasting the evolution equations as autonomous systems, one can identify critical points corresponding to cosmological solutions (e.g., matter domination, de Sitter expansion) and analyze their stability. Previous studies considered DBI fields with inverse square \citep{copeland1}, exponential \citep{guoexp}, and hyperbolic potentials \citep{Quiros}, revealing rich phase space structures. In particular, Silverstein and Tong \citep{tong} showed that D3 brain motion in warped throats generates generalized DBI actions with non-canonical kinetic terms, offering new phenomenological possibilities.

In the context of cosmological models that address the acceleration of the universe, scalar fields play an integral role. Specifically, the D-brane inflation (DBI) model has garnered considerable attention due to its ability to provide a self-consistent framework for scalar field dynamics in a high-energy regime. The DBI scalar field, when coupled with various types of potentials, can describe a wide range of cosmological phenomena, including inflationary dynamics and DE.
  
One of the main troubles of using string theory in cosmology directly is the so-called no-go theorem \citep{hao,chingangbam}, for wrapped products by compactifying the extra dimensions. In string theory, it was predicted by Sen \citep{sen1,sen2,sen3} that there are 
tachyon fields in both open and closed string theory. For more on open and closed string theory, one can refer to \citep{polchinski}.  Even though for closed string theory, the tachyon fields are projected out in open string, they remain even though one can use a spontaneous symmetry-breaking argument to get rid of tachyon modes, one can still fully explain the reason for its existence. In bosonic string theory, if one uses the Nambu-Goto action, then it is almost impossible to quantize in order to get meaningful quantization rules. Therefore, we instead introduce the conformally invariant Polyakov action. Using conformal field theory techniques, this action can be quantized, although the resulting spectrum contains undesirable tachyonic modes. Although such modes violate causality, they can be shown to be unstable. So, tachyon modes are typically given by the DBI Lagrangian density, which has the following form,
\begin{equation}\label{4a}
\mathcal{L}_\mathrm{DBI}=-V(\phi)\sqrt{1+\partial \phi^2},
\end{equation}
where $V(\phi)$ is a potential function for the scalar field and  $\partial\phi^2=\partial^{\mu}\phi\partial_{\mu}\phi$ denotes the kinetic term for the tachyon fields. We would like to note that the form written above is often mentioned as a tachyon field, as DBI-like fields are often defined with a minus constant sign in order to be consistent with the ordinary scalar field Lagrangian. However, in our case, since we are working with the equation of motion of the field and the energy-momentum tensor, the overall constant does not matter after the variation, so we have adopted the term "DBI field" as it appears in the literature rather than "tachyon form for DBI."

From the Lagrangian, one can find the field equation for the tachyon field from the Euler-Lagrangian equation as
\begin{equation}\label{4b}
    \frac{\Ddot{\phi}}{1-\dot{\phi}^2}+3H\dot{\phi}+\frac{V_{,\phi}}{V}=0,
\end{equation}
which is called the modified Klein-Gordon equation for the DBI field.\\
For the full Lagrangian density becomes
\begin{equation}
    \mathcal{L}_\mathrm{tot}= \frac{1}{2\kappa}R+\mathcal{L}_\mathrm{DBI},
\end{equation}
where $\kappa=\frac{8\pi G}{c^4}$, in natural units we take $8\pi G=c=1$, so we are taking $\kappa=1$
so the Einstein-Hilbert action becomes
\begin{equation}
    S= \int \left[\frac{1}{2} R + \mathcal{L}_\mathrm{DBI}\right] \sqrt{-g} d^4x.
\end{equation}
Now, one can vary this action with respect to the metric $g_{\mu\nu}$ and recover Einstein's field equations;
\begin{equation}
    G_{\mu\nu}= R_{\mu\nu}-\frac{1}{2}R g_{\mu\nu}= T_{\mu\nu}.
\end{equation}
If one puts the flat FLRW metric, which is given as
\begin{equation}
    ds^2= -dt^2+a(t)^2(dr^2+r^2d\theta^2+r^2sin^2\theta d\phi^2).
\end{equation}
One recovers the following Friedman equations given as,
\begin{equation}\label{4c}
    3H^2= \rho_{\phi} + \rho_\mathrm{m},
\end{equation}
\begin{equation}\label{4d}
    \dot{H}=-\frac{1}{2} [\rho_{\phi} + p_{\phi}+\rho_\mathrm{m}+p_\mathrm{m}].
\end{equation}
The energy density $\rho_{\phi}$ and pressure $p_{\phi}$ for the DBI scalar field are given by
\begin{equation}\label{4e}
    \rho_{\phi}=\frac{V}{\sqrt{1-\dot{\phi}^2}},
\end{equation}
\begin{equation}\label{4f}
    p_{\phi}=-V\sqrt{1-\dot{\phi}^2},
\end{equation}
respectively. In addition, the corresponding equation of the state parameter $w_{\phi}$ and the density parameter $\Omega_{\phi}$ for the DBI scalar field are defined as
\begin{equation}\label{w}
    w_{\phi}=\frac{p_{\phi}}{\rho_{\phi}}=\dot{\phi}^2-1,
\end{equation}
\begin{equation}\label{density}
    \Omega_{\phi}=\frac{V}{3H^2\,\sqrt{1-\dot{\phi}^2}},
\end{equation}
respectively.

Further, by rewriting the Friedmann equations in the presence of a DBI scalar field and pressureless matter ($p_\mathrm{m}=0$), we obtain the following result.
\begin{equation}
\label{RFE1}
    3H^2-\rho_\mathrm{m}=\rho_{\phi}=\frac{V}{\sqrt{1-\dot{\phi}^2}},
\end{equation}
\begin{equation}
\label{RFE2}
    2\dot{H}+3H^2=-p_{\phi}-p_\mathrm{m}=-p_{\phi}=V \sqrt{1-\dot{\phi}^2}.
\end{equation}
Using Eqs. \eqref{RFE1} and \eqref{RFE2}, the scalar field potential can be expressed as
\begin{equation}\label{V}
    V^2=(2\dot{H}+3H^2)(3H^2-\rho_\mathrm{m}),
\end{equation}
and the kinetic term satisfies the following relation:
\begin{equation}\label{KE}
    (1-\dot{\phi}^2)=\frac{(2\dot{H}+3H^2)}{(3H^2-\rho_\mathrm{m})}.
\end{equation}

For the case of pressureless matter $p_\mathrm{m}=0$, the matter density is given by $\rho_m(z)=3H_0^2\Omega_\mathrm{m,0}(1+z)^3$, where the subscript zero refers to the values measured at the present epoch. Here, $z$ represents the redshift, which is defined as $z=\frac{1}{a}-1$, with $a(t)$ being the scale factor. 

In this study, we need to express the time dependence of the equation in terms of redshift $z$ to investigate the observational study. This can be achieved using the relation
\begin{equation}
\label{relation}
    \frac{d}{dt}=-(1+z)H(z)\frac{d}{dz}.
\end{equation}

\section{Model--independent Reconstructions}\label{sec4}
As we discussed in the previous section, the DBI field offers a consistent and physically motivated description of late-time dark energy. In this work, we therefore attempt to reconstruct the corresponding DBI dynamics directly from late-time data using advanced, model-independent techniques such as the Gaussian Process algorithm.
\subsection{Observational Data}
For the GP reconstruction, we employ two datasets. The first consists of 58 observational Hubble data (OHD) points, comprising 26 points from radial BAO measurements and 32 points from Cosmic Chronometers (CC). The OHD is taken in the redshift interval $0.07 \le z \le 2.42$. The second dataset includes 5 recently released data points from the DE Spectroscopic Instrument (DESI).

  The CC data consist of 32 measurements derived from the differential age evolution of passively evolving galaxies, which yield $H(z)$ in a model-independent fashion \citep{Jimenez2002, Moresco2016}. The rest comes from Radial BAO data, which consists of 26 measurements based on the position of the baryon acoustic oscillation peak as a standard ruler, calibrated by the sound horizon scale \citep{Gaztanaga2009, Alam2017}. It is worth mentioning that the CC data are taken purely on the basis of the distance ladder, so no assumptions have been made regarding the cosmological models, but it is also unable to give any constraints on the $H(z)$ function. However, for the Radial BAO data, one has to assume the position of the sound horizon in order to get the pick position. So, this makes the 26 Radial BAO data model dependent. So, our combined OHD data sets break the degeneracy of the model dependency by noting that CC data is purely based on the distance ladder. 
  When converting radial BAO measurements to $H(z)$, we adopt the fiducial sound horizon $r_d = 147.09 \pm 0.26$~Mpc from Planck 2018 $\Lambda$CDM~\cite{P18}, following standard practice in late-time analyses. While a fully empirical $r_d$ calibration using CC data alone is an interesting direction, current dataset coverage limits its precision relative to early-universe determinations; we therefore retain the Planck fiducial value here and plan to explore data-driven $r_d$ inference in future work with improved $H(z)$ measurements.
  The full 58 data points with proper citation can be found in the paper by G. Gadbail et al. \citep{gaurav1}.
It is worth mentioning that Cosmic Chronometer (CC) data provide direct, model-independent measurements of the Hubble parameter $H(z)$ by utilizing the differential ages of passively evolving galaxies. Unlike distance-based probes, CC measurements do not rely on a distance ladder or an assumed cosmological model, making them particularly valuable for non-parametric reconstructions of the expansion history. 
  
  We also used the five data points from the latest DESI survey \citep{desi,desi1}. The DE Spectroscopic Instrument (DESI) employs a multi-tracer approach, which includes 1. the Galaxy Survey (Including Bright Galaxy Survey (BGS), Luminous Red Galaxies (LRGs), and Emission Line Galaxies (ELGs), covering redshifts up to 1.6) 2. Quasars (Distant, luminous objects with redshifts up to 2.1) and finally 3. Lyman-$\alpha$ Forest: The absorption features in the spectra of quasars caused by intervening hydrogen clouds, allowing measurements up to redshift 2.33.
  
  By analyzing more than 5.7 million objects across a 7500 square degree area, DESI has achieved a combined precision of approximately 0.52 percent in BAO measurements. The highest significance of BAO detection is 9.1$\sigma$ at an effective redshift of 0.93, with a constraint of 0.86 percent placed on the BAO scale. One can also note that at redshift 2.33, DESI measures a transverse comoving distance $(D_L)$ and Hubble parameter $(H_0(z))$ at redshift 2.33. These make it an ideal candidate for the distance ladder, as two different and somewhat independent observations of the cosmological parameters break the degeneracy in the measurement. The DESI collaboration has recently released high-precision measurements of the baryon acoustic oscillation (BAO) scale and redshift-space distortions, providing constraints on the comoving angular diameter distance $D_M(z)$ and the Hubble parameter $H(z)$ across multiple redshift bins (e.g., $z \approx 0.1$--$2.3$). These geometric observables are derived from the galaxy and quasar clustering power spectra and are reported with full covariance matrices that properly account for the intrinsic correlations between $D_M(z)$ and $H(z)$. Unlike luminosity distance $D_L(z)$, which is directly probed by standard candles, DESI's BAO measurements offer complementary, model-agnostic constraints on the cosmic expansion history.
  
  CC data provide direct, model-independent measurements of $H(z)$, but they are sparse and cannot provide useful constraints on $f(z)$. BAO data, although model-dependent, offer high precision and improve coverage across redshifts. Therefore, combining CC with BAO and DESI increases statistics, reduces uncertainties, and leads to a more stable and accurate GP reconstruction of $H(z)$.
\subsection{Gaussian Process}
$\,\,\,\,\,$Gaussian Processes (GPs) are widely used in machine learning as a non-parametric Bayesian method for reconstructing continuous functions and their derivatives directly from noisy data \citep{Rasmussen2005}. The key strength of GP lies in its model-independent nature: instead of assuming a specific functional form to fit the data, GP infers the function by placing a prior over functions, guided by a covariance kernel. This makes it particularly suitable for cosmology, where one often aims to reconstruct quantities such as the Hubble parameter $H(z)$, its derivatives, or the DE equation of state without assuming any specific cosmological model \citep{Seikel2012, Seikel2013, Mehrabi2021}.

In this framework, the unknown function $f(x) \equiv H(x)$ is modeled as a GP
$$
f(x)\sim\mathcal{GP}\bigl(m(x),\,k(x,x')\bigr),
$$
where $m(x) = \mathbb{E}[f(x)]$ is the prior mean function, often chosen as a constant or low-order polynomial, and $k(x,x')=\text{cov}(f(x),f(x'))$ is the kernel function that determines the covariance between observations at points $x$ and $x'$. There are a lot of covariance functions that we can choose \citep{Rasmussen2005,Seikel2013}. Since the Hubble parameter is expected to evolve smoothly with redshift, it is natural to choose a covariance function that depends only on the separation between data points. For this reason, and following its successful use in similar cosmological reconstruction studies \citep{Seikel2012,Busti2014,sun/2021, Yahya:2013xma,Zhang:2023pis}, we adopt the squared exponential covariance function, which is given by
$$
k(x,x') = \sigma_f^2\exp\!\left[-\frac{(x - x')^2}{2\ell^2}\right].
$$
This covariance function depends on two hyperparameters, $\ell$ and $\sigma_f^2$, which encode the smoothness of the reconstructed function through a characteristic length scale and the output variance, respectively. This approach enables a robust, data-driven reconstruction of cosmological quantities, making GP a powerful tool for probing the expansion history of the universe and the nature of DE.

Importantly, derivatives of $H(z)$ are obtained analytically by differentiating the kernel
$$
\frac{\mathrm{d}^m}{\mathrm{d}x^m}\frac{\mathrm{d}^n}{\mathrm{d}x'^n}k(x,x').
$$
Since the kernel maintains its Gaussian form even after taking derivatives, we can easily reconstruct $H'(z)$ and $H''(z)$, together with their uncertainties \citep{Seikel2013}. This feature is crucial for diagnostics of cosmic acceleration and model selection. It is important to note that the GP reconstruction of $H(z)$ and $H'(z)$ provides a smooth, noise-reduced representation of the expansion history that is fully consistent with the observational data. The pointwise $1\sigma$ uncertainties $\sigma_H$ and $\sigma_{H'}$ obtained from the GP are propagated consistently through all subsequent reconstructions performed in this work, ensuring that the error budget at each step faithfully reflects the uncertainties present in the underlying observational dataset.

\begin{figure*}[htbp]
\centering

\begin{minipage}[b]{0.42\textwidth}
  \centering
  \includegraphics[width=\linewidth]{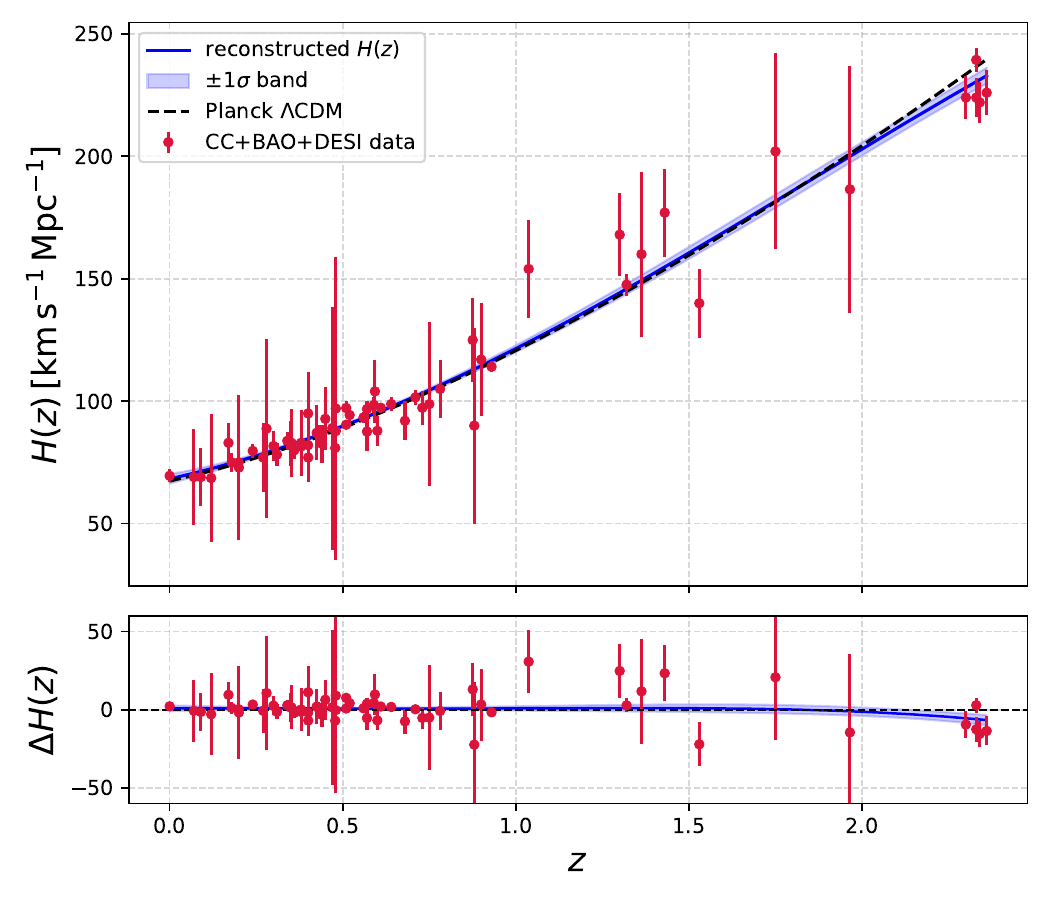}
\end{minipage}\hfill%
\begin{minipage}[b]{0.54\textwidth}
  \centering
  \includegraphics[width=\linewidth]{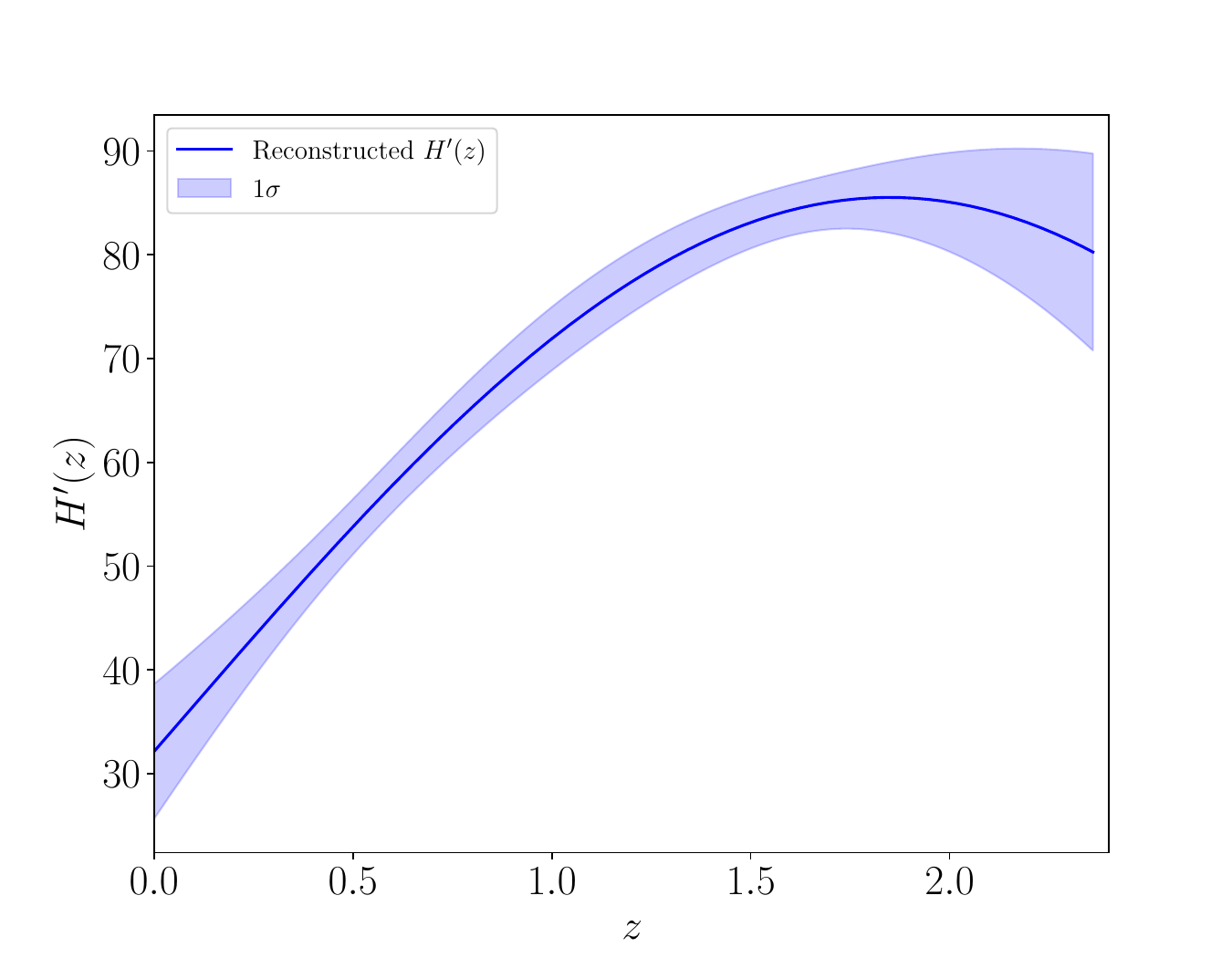}
\end{minipage}

\caption{\justifying In the left panel, we present the reconstructed $H(z)$ using 32 CC, 26 radial BAO, and 5 DESI data points; the red solid points with error bars represent the observational measurements. In the right panel, we show the first-order derivative of $H(z)$ with respect to $z$. In both plots, the solid blue line corresponds to the mean GP reconstruction, while the shaded light blue region denotes the $1\sigma$ uncertainty.}
\label{fig:01}
\end{figure*}

Finally, we implement this methodology using the publicly available \texttt{GaPP} code \citep{Seikel2012}, which efficiently handles matrix inversions ($\mathcal{O}(N^3)$) and hyperparameter optimization to reconstruct $H(z)$ and its redshift derivative from the combined CC+BAO+DESI data. The results, along with the corresponding $1\sigma$ confidence bands, are presented in Fig.~\ref{fig:01}. 

\subsection{Posterior Sampling of $H_0$}
$\,\,\,\,\,$The GP reconstruction provides, at each redshift point $z_i$, a mean value $H_{\text{rec}}(z_i)$ together with a corresponding $1\sigma$ uncertainty $\sigma_{\text{rec}}(z_i)$. This can be viewed as a Gaussian distribution, 
\begin{equation}
    H(z_i) \sim \mathcal{N}(H_{\text{rec}}(z_i), \sigma_{\text{rec}}^2(z_i)),
\end{equation}
representing the range of possible realizations of the Hubble parameter consistent with both the observational data and the GP prior. To propagate this uncertainty into an estimate of the Hubble constant, we employed a Monte Carlo sampling procedure. For each redshift $z_i$, we generate $10^4$ random samples from the Gaussian distribution defined above, thereby producing an ensemble of realizations of the reconstructed expansion history $H(z)$. Flattening these realizations across all redshift points yields the posterior probability distribution for the Hubble parameter, and, in particular, evaluating this distribution at $z=0$ corresponds to the posterior of the Hubble constant, $H_0$. Finally, we fitted this posterior with a Gaussian profile to obtain its mean and standard deviation, ensuring that the uncertainties from the GP reconstruction were faithfully propagated into the final constraint.


\begin{figure}
    \centering
    \includegraphics[width=1\linewidth]{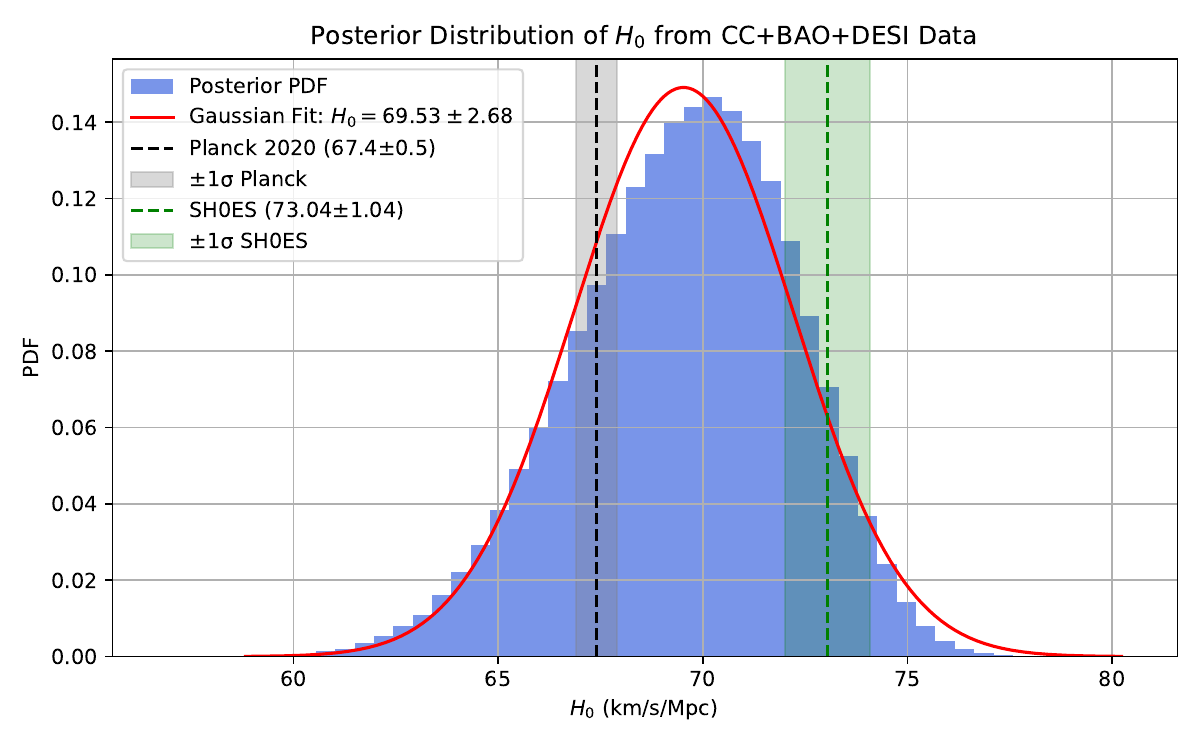}
    \caption{\justifying The posterior probability distribution of the Hubble constant $H_0$ reconstructed from the combined CC + BAO + DESI dataset. The histogram represents the Monte Carlo realizations of $H_0$, while the solid curve denotes the best-fit Gaussian with $H_0 = 69.53 \pm 2.68$ km s$^{-1}$ Mpc$^{-1}$. The vertical shaded regions correspond to the $1\sigma$ intervals from the Planck 2020 ($67.4 \pm 0.5$) and SH0ES ($73.04 \pm 1.04$) measurements, respectively, shown for comparison.}
    \label{fig:02}
\end{figure}

The resulting posterior distributions of $H_0$ yield $H_0 = 69.53 \pm 2.68$ km s$^{-1}$ Mpc$^{-1}$. This result is illustrated in Fig.~\ref{fig:02}, where the histograms of the Monte Carlo realizations of $H_0$ are shown together with their best-fit Gaussian curves. 

It is worth noting that $H_0$ errors inferred from GP reconstruction are kernel-dependent~\cite{Eoin/2021}. The squared exponential kernel ($\nu \to \infty$), adopted in this work, is the most aggressive choice in the sense that it yields the smallest errors on $H_0$ among the Mat\'ern kernel class. We observe this effect in the kernel sensitivity analysis presented by \cite{Gadbailprd} in their recent work on model-independent $H_0$ measurements, where they demonstrate that Hubble constant estimates derived using Matern kernels with smoothness parameters $\nu = 5/2, 7/2, 9/2$ exhibit significant sensitivity to kernel choice, with $H_0$ values shifting by approximately $8 \, \text{km s}^{-1} \text{Mpc}^{-1}$ across different kernels for the same dataset combinations. This follows from its infinite differentiability, which enforces the strongest correlations between $H(z=0)$ and $H(z_i)$, thereby most tightly constraining $H_0$. Kernels with finite $\nu$ would produce larger $H_0$ errors. Our quoted uncertainties on $H_0$ should therefore be understood as lower bounds on the true GP uncertainty.
\subsection{Reconstructed $w_{\phi}$ and $\Omega_{\phi}$}
In this subsection, we reconstructed the DE EoS parameter $w_{\phi}$ and the DE density parameter $\Omega_{\phi}$ using the above reconstructed Hubble function and its derivative.

Using the value of the matter density parameter and Eq. \eqref{KE} in Eqs. \eqref{w} and \eqref{density}, we get the simplified $w_{\phi}$ and $\Omega_{\phi}$ as
\begin{equation}\label{ww}
    w_{\phi}=\frac{-2(1+z)H(z)\frac{dH}{dz}+3H^2}{3H^2-3H_0^2\,\Omega_\mathrm{m,0}\,(1+z)^3},
\end{equation}
and 
\begin{equation}
    \Omega_{\phi}=1-\frac{H_0^2\,\Omega_\mathrm{m,0}\,(1+z)^3}{H^2}.
\end{equation}

\begin{figure*}
\centering
\includegraphics[width=0.47\textwidth]{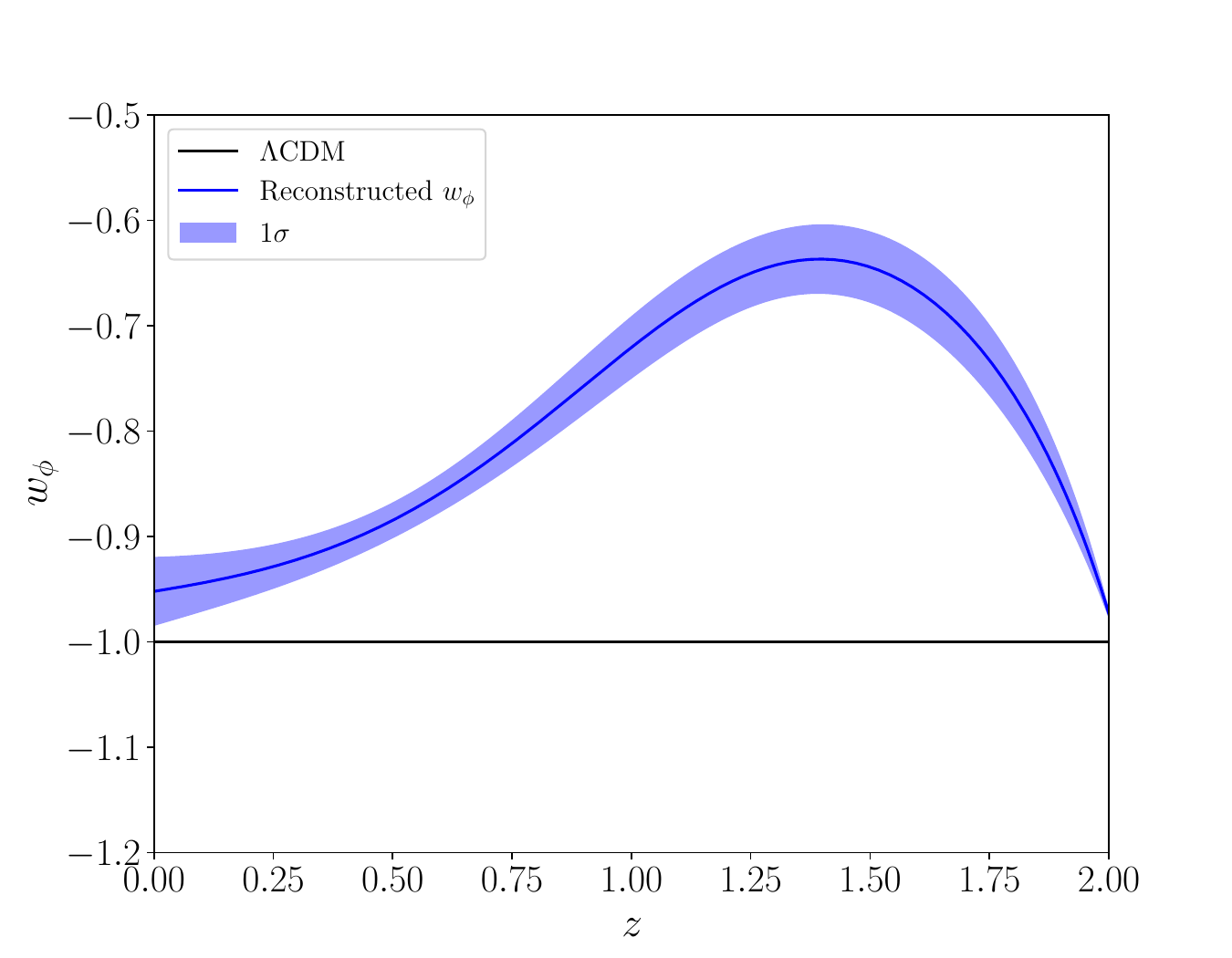}
 \hspace{0.15in} 
\includegraphics[width=0.47\textwidth]{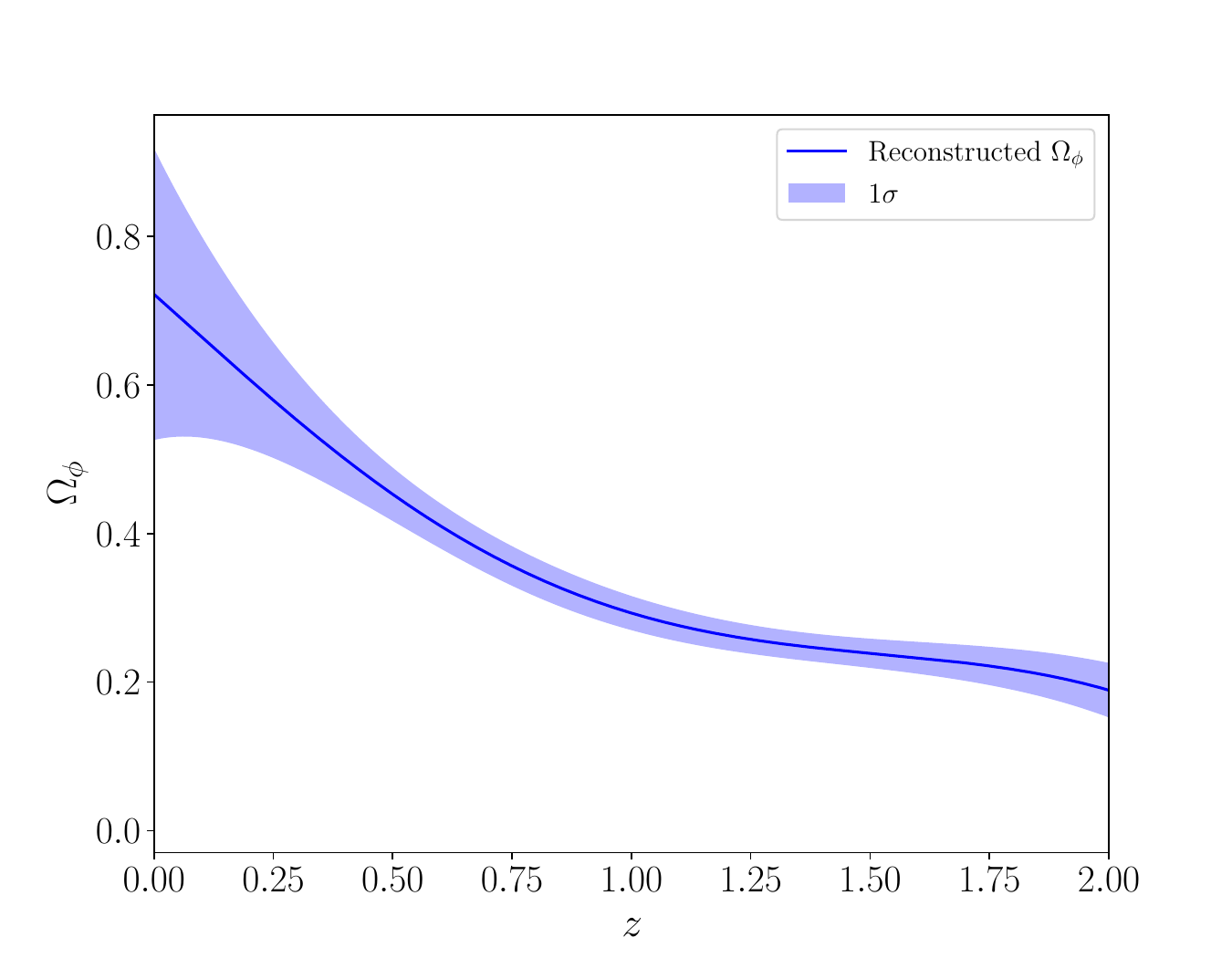}
\\ 
\caption{\justifying In the left panel, we have plotted the reconstructed mean equation of state $w_\phi$ (the blue line) using 32 CC + 26 Radial BAO + 5 DESI data points; the light blue region shows the $1\sigma$ uncertainty associated with the GP. We have also plotted the $\Lambda$CDM EoS, which is -1 for comparison purposes. In the right panel, we have reconstructed the mean total DE density $\Omega_\phi$ (the blue line), with the light blue region showing the $1\sigma$ uncertainty associated with the GP.}
\label{fig:03}
\end{figure*}

To plot the above quantities, we utilized the reconstructed functions $H(z)$ and $H'(z)$, together with the prior value of $H_0$ obtained from our GP analysis, while $\Omega_{\mathrm{m},0} = 0.279$ is assumed as a fiducial value.
We adopt the squared exponential kernel because its infinite differentiability ensures stable reconstruction of $H'(z)$, whereas Matern kernels amplify numerical noise in the derivative and propagate unphysically large uncertainties into $w_\phi(z)$ and $V(\phi)$~\cite{Seikel2012,GomezValent2018}.
The resulting reconstructions of $w_{\phi}(z)$ and $\Omega_{\phi}(z)$ with 1$\sigma$ error are presented in Fig.~\ref{fig:03}. From the analysis, we obtain the current value of the DE equation of state as $w_{\phi,0} = -0.9521 \pm 0.0109$, and the present value of the DE density parameter as $\Omega_{\phi,0} = 0.7220 \pm 0.3923$. These results agree well with the standard $\Lambda$CDM model, where $w_{\Lambda,0} = -1$ and $\Omega_{\Lambda,0} \approx 0.7$, indicating the consistency of our reconstruction approach with the current cosmological paradigm. Even though, as we have seen from Eq. \eqref{w}, in the slow-roll approximation, that is, $\dot{\phi} \ll 1$, $w_{\phi}$ always becomes $w_{\phi} \approx -1$. What our analysis shows is how fast it reaches $-1$ without invoking any prior model. 
We note that $w_\phi(z)$ depends on $H'(z)$, whose uncertainty is amplified by numerical differentiation~\cite{Holsclaw2010a}; the bands in Fig.~3 propagate GP covariances via Monte Carlo and represent lower bounds within the squared-exponential prior, warranting conservative interpretation of dynamical dark energy claims.

The deviations visible in Fig.~\ref{fig:03} therefore reflect the theoretical effect of the DBI framework rather than an inconsistency in the data. Also, it is important to see that the EoS $w_\phi$ in Eq.~\eqref{ww} is specific to the DBI scalar field and is structurally distinct from the standard fluid-based EoS. Furthermore, we note that the EoS involves $H'(z) = dH/dz$, whose relative uncertainty from the GP reconstruction is substantially larger than that of $H(z)$ itself, since numerical differentiation amplifies noise. Hence, within the DBI framework, $w_\phi$ naturally exhibits a deviation from $-1$ compared to the reconstructed $H(z)$ shown in Fig.~\ref{fig:01}, and this behavior is a direct consequence of the non-canonical dynamics of the DBI scalar field.

Moreover, the DE EoS parameter for the DBI scalar field crosses the phantom divide and becomes less than $-1$ for $z > 2.1$. This behavior implies that $\dot{\phi}^2 < 0$, rendering the scalar field $\phi$ imaginary, which is unphysical within the framework of the proposed model. However, a fundamental physical requirement of this model is that the kinetic term must remain non-negative ($\dot{\phi}^2 \ge 0$), which consequently ensures $w \ge -1$ at all times. Therefore, we conclude that the DBI scalar field model is consistent with the CC+BAO+DESI data within the redshift interval $0<z\leq 2.1$, beyond which the reconstruction enters a region that is not physically viable in this framework. Furthermore, to check the effect on the EoS phantom divide crossing line, 
we check the Matern kernels with the smoothness parameter $\nu = 5/2$ in 
our analysis. Our independent examination confirms that the adoption of 
this alternative kernel (Matern with $\nu = 5/2$) results in substantially 
larger uncertainty bands compared to our fiducial squared exponential choice, 
and the equation of state exhibits a phantom divide crossing at $z\approx 2.0$ 
rather than $z > 2.1$. The phantom divide crossing location is sensitive to 
covariance function choice, but the qualitative conclusion regarding DBI 
model inconsistency at $z \gtrsim 2$ persists across kernels, albeit with 
appropriately enlarged confidence intervals that reflect this methodological 
uncertainty.

\section{Reconstruction of DE Scalar Field Potential}\label{sec5}
In this section, we reconstruct the DE scalar field potential using the GP framework described above. To perform this reconstruction, we first outline the methodological setup; subsequently, we apply the chi-square fitting method for model selection; and finally, we constrain the model parameters using the Markov Chain Monte Carlo (MCMC) technique as detailed in the following subsections.
\subsection{Method Setup}
Now, using the relation \eqref{relation} and the value of matter density $\rho_\mathrm{m}(z)$, we rewrite the scalar field potential (Eq. \eqref{V}) in terms of redshift $z$ as
\begin{multline}\label{V(z)}
    V^2(z)=-6\,H(1+z)\frac{dH}{dz}\left(H^2-H_0^2\,\Omega_\mathrm{m,0}\,(1+z)^3\right)+\\
    9H^2\left(H^2-H_0^2\,\Omega_\mathrm{m,0}\,(1+z)^3\right),
\end{multline}
and the kinetic term of the scalar field (Eq. \eqref{KE}) becomes 
\begin{equation}
    \dot{\phi}^2(z)=\frac{-2H(1+z)\frac{dH}{dz}+3H_0^2\Omega_\mathrm{m}\,(1+z)^3}{3H_0^2\Omega_\mathrm{m}\,(1+z)^3-3H^2}.
\end{equation}
Using relation~\eqref{relation}, the kinetic term can be rewritten as
\begin{align}
 \frac{d\phi}{dz}&=\frac{1}{(1+z)\,H}\left(\frac{-2H(1+z)\frac{dH}{dz}+3H_0^2\Omega_\mathrm{m,0}(1+z)^3}{3H_0^2\Omega_\mathrm{m,0}(1+z)^3-3H^2}\right)^{\frac{1}{2}}.
 \end{align}
 
Since an analytic solution to the above differential equation is difficult to obtain, we proceed by seeking a numerical solution. To facilitate numerical integration, we derive a recursive relation between consecutive redshift points $z_i$ and $z_{i+1}$. This allows us to express $\phi(z_{i+1})$ in terms of $\phi(z_i)$, along with the values of $H(z_i)$ and $H'(z_i)$, as
 \begin{multline}\label{Recc}
   \phi(z_{i+1})=\phi(z_i)+\frac{(z_{i+1}-z_i)}{(1+z_i)\,H(z_i)}\\\left(\frac{-2H(1+z_i)\frac{dH}{dz}\big|_{z_i}+3H_0^2\,\Omega_\mathrm{m,0}(1+z_i)^3}{3H_0^2\,\Omega_\mathrm{m,0}(1+z_i)^3-3H(z_i)^2}\right)^{\frac{1}{2}},
\end{multline}
where
$$
\phi'(z) = \frac{\phi(z + \Delta z) - \phi(z)}{\Delta z},
$$
for small $\Delta z$.
We emphasize that the reconstruction of $V(\phi)$ via Eqs.~(21)--(24) is conditional on the DBI scalar field framework; different dynamical models can reproduce the same $H(z)$, so the inferred potential is not unique but represents the DBI-consistent form compatible with the data~\cite{Zhao2017}.
Furthermore, we emphasize that the apparent redshift evolution of $w_\phi(z)$ is indicative rather than conclusive; distance-based measurements have intrinsically limited sensitivity to dark energy dynamics, and the tendency toward $w=-1$ at higher $z$ may partly reflect the GP prior reverting to smoothness~\cite{Holsclaw2010a}.
We further note that the reconstruction is data-driven for $z \lesssim 1.5$, where CC+BAO+DESI coverage is sufficient; at $z \gtrsim 1.5$, the GP increasingly reverts to the squared-exponential prior, so the approach of $w_\phi(z) \to -1$ should not be interpreted as a physically meaningful detection~\cite{Seikel2012}.
According to Eq.~\eqref{Recc}, in order to determine the values of $\phi(z)$, it is necessary to know the values of $H(z)$, $H'(z)$, $H_0$, and $\Omega_\mathrm{m,0}$. To obtain $H(z)$ and $H'(z)$ in the redshift range $z \in [0,\,2.5]$, we employed the GP reconstruction method using the combined CC+BAO+DESI dataset. In our analysis, we adopted the prior value $H_0 = 69.53 \pm 2.68 , \text{km s}^{-1} \text{Mpc}^{-1}$ obtained from our GP reconstruction, while $\Omega_{\mathrm{m},0} = 0.279$ is assumed as a fiducial value.

We emphasize that the reconstruction of $\phi(z)$ and subsequently $V(\phi)$ via Eq.~\eqref{Recc} is conditional on several assumptions. First, the GP reconstruction of $H(z)$ and $H'(z)$ adopts the squared-exponential covariance kernel, which imposes smoothness and infinite differentiability on the inferred expansion history. Second, the conversion of BAO measurements to $H(z)$ relies on the fiducial sound horizon scale $r_d = 147.09 \pm 0.26$ Mpc, consistent with Planck 2018 $\Lambda$CDM. Third, the mapping to the scalar field quantities explicitly assumes the DBI Lagrangian structure, a flat FLRW metric, and the fiducial matter density $\Omega_{\mathrm{m},0} = 0.279$. Therefore, while our approach is non-parametric with respect to the analytical form of $V(\phi)$, the physical interpretation of the reconstructed potential is inherently tied to the DBI framework and the aforementioned priors.

\subsection{Model Selection and Chi-square Fitting}
\begin{figure*}[htbp]
\centering

\begin{minipage}{0.45\textwidth}
  \centering
  \includegraphics[width=\linewidth]{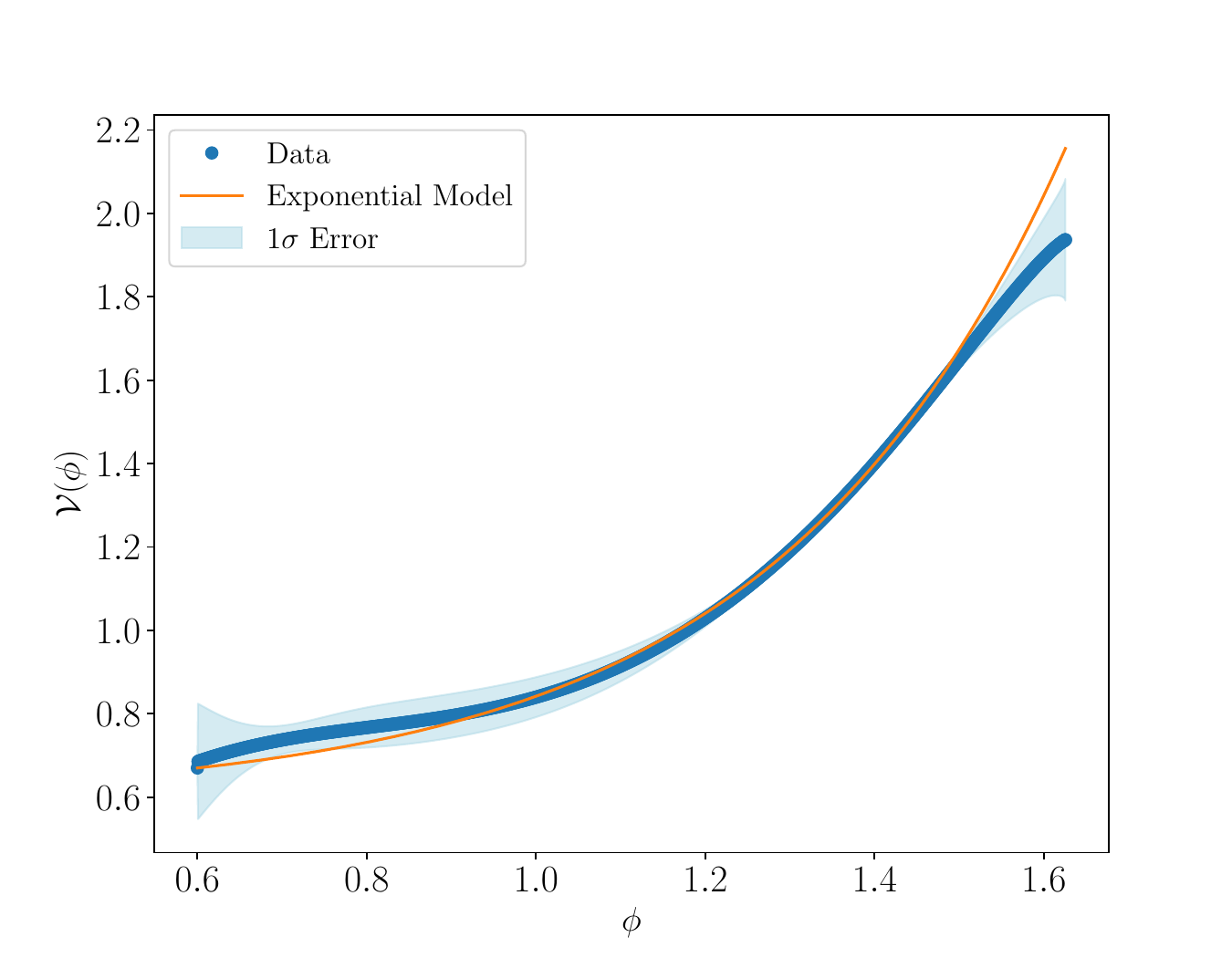}
  \vskip1ex
  \small\textbf{(a) Exponential Potential}
\end{minipage}\hspace{0.05\textwidth}%
\begin{minipage}{0.45\textwidth}
  \centering
  \includegraphics[width=\linewidth]{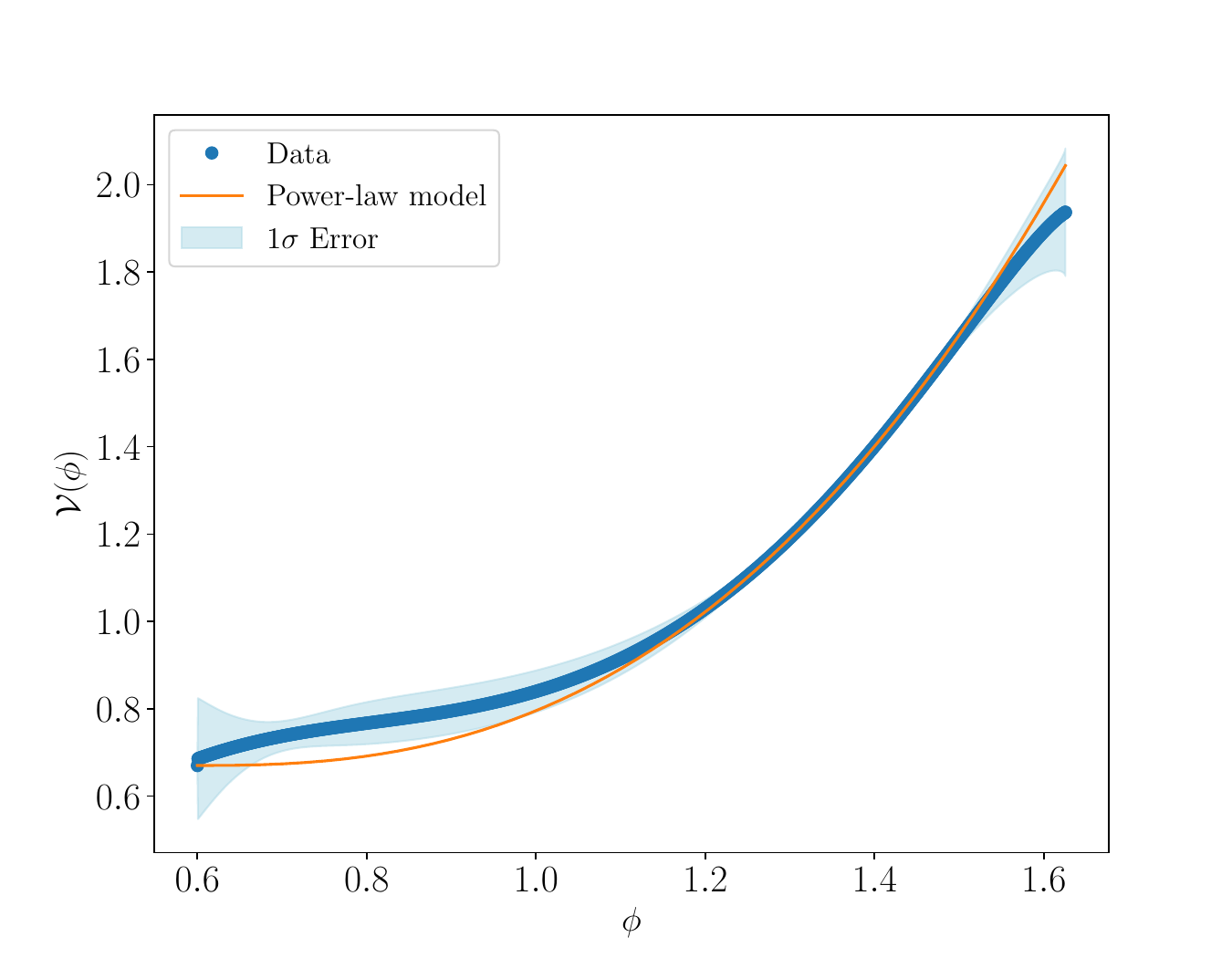}
  \vskip1ex
  \small\textbf{(b) Power-law Potential}
\end{minipage}

\vspace{0.6cm}

\begin{minipage}{0.45\textwidth}
  \centering
  \includegraphics[width=\linewidth]{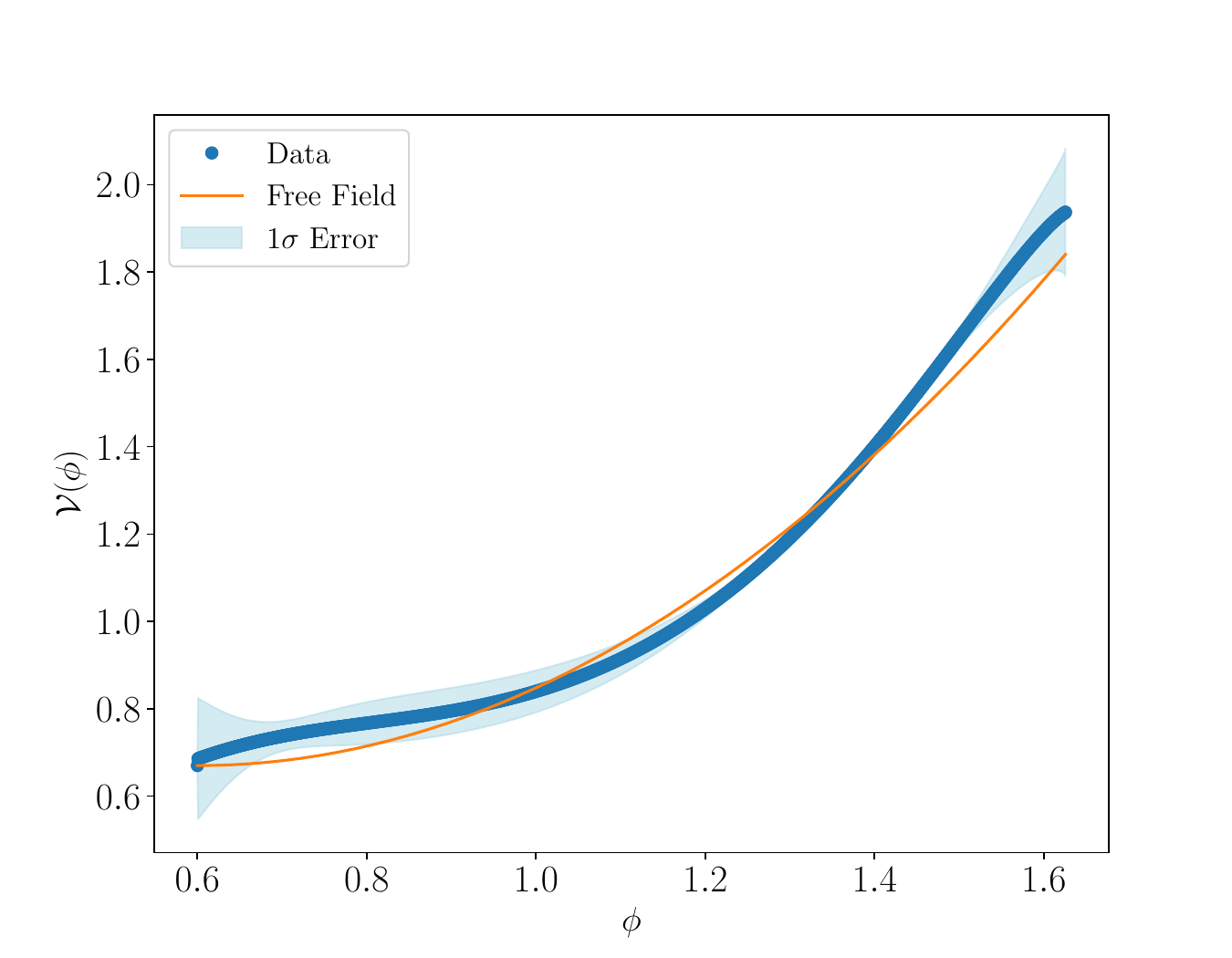}
  \vskip1ex
  \small\textbf{(c) Free Field Potential}
\end{minipage}\hspace{0.05\textwidth}%
\begin{minipage}{0.45\textwidth}
  \centering
  \includegraphics[width=\linewidth]{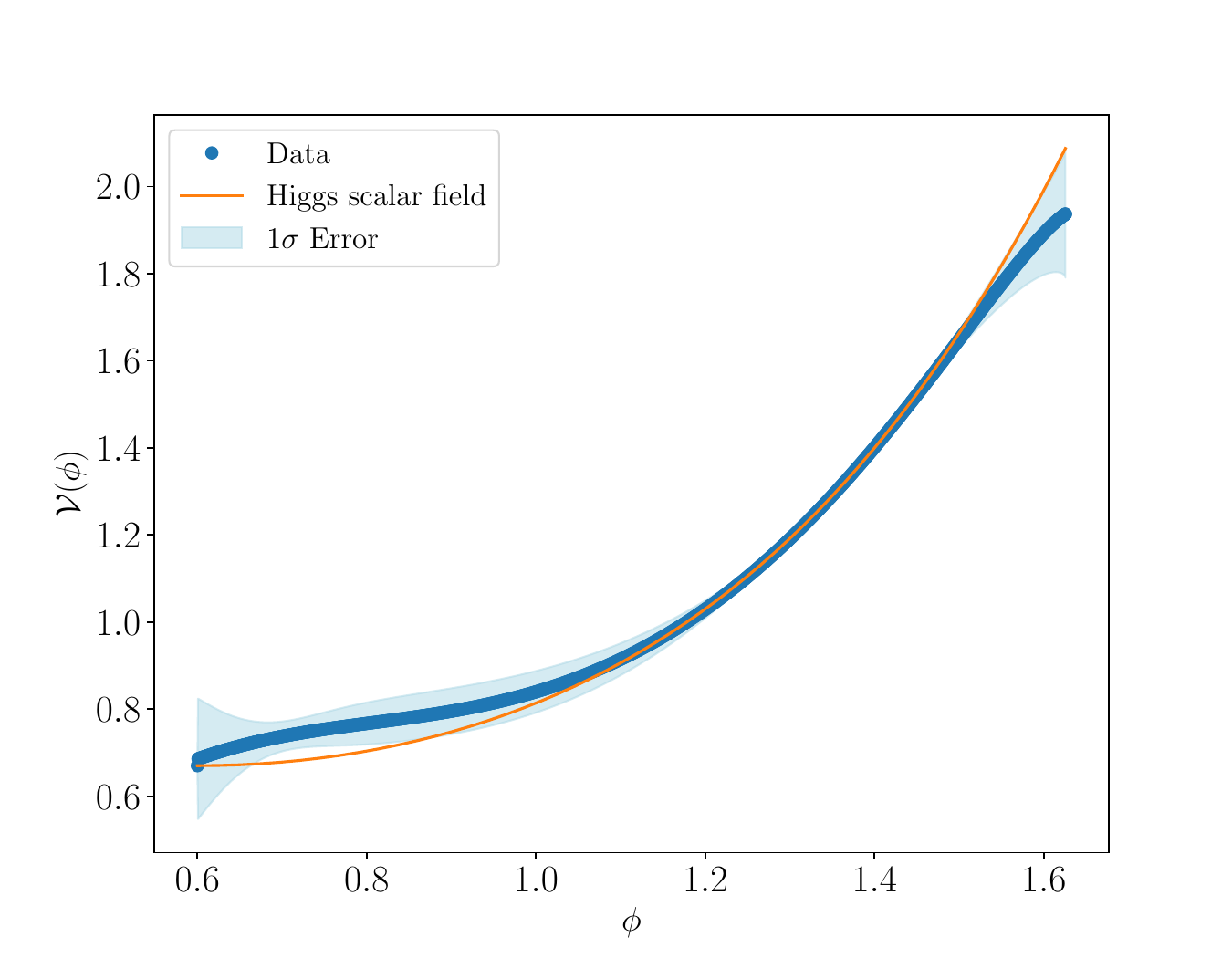}
  \vskip1ex
  \small\textbf{(d) Higgs-like Potential}
\end{minipage}

\caption{\justifying This figure shows the reconstructed scalar field potential with 1$\sigma$ uncertainty using the CC+BAO+DESI datasets via a GP. Standard theoretical scalar field potential models are fitted to the reconstructed data using chi-square curve fitting. The orange line represents the theoretical model, while the thick blue dots with error bars (light blue) denote the reconstructed data with 1$\sigma$ uncertainty.}
\label{fig:04}
\end{figure*}

Using Eqs.~\eqref{V(z)} and \eqref{Recc}, we reconstruct the dimensionless scalar field potential $\mathcal{V}(\phi)=V/3H_0^2$ of DE and obtain the dataset in the form of $\mathcal{V}(\phi)$ versus $\phi$, along with its associated uncertainties $\sigma_{\mathcal{V}}$. In Fig. \ref{fig:04}, we can see these reconstructed data of scalar field potential (blue dots) with $1\sigma$ error represented by light blue. The uncertainties $\sigma_{\mathcal{V}}$ on the reconstructed potential $\mathcal{V}(\phi)$ are propagated from the GP output 
$\sigma_H$ and $\sigma_{H'}$ through the analytical expressions in  Eqs.~\eqref{V(z)} and \eqref{Recc} at each redshift point. This ensures that the reconstructed dataset $\{\mathcal{V}(\phi_i) \pm \sigma_{\mathcal{V}}(\phi_i)\}$ consistently carries the full observational uncertainty from the original 
CC+BAO+DESI data through the GP reconstruction.

  To identify the most suitable model corresponding to our reconstructed potential, we performed a chi-square fitting analysis using standard theoretical scalar field models. The chi-square statistic is given by
\begin{equation}
\chi^2 = \sum \frac{\left(\mathcal{V}_{SM}(\theta_s,\phi_i) - \mathcal{V}_{obs}(\phi_i)\right)^2}{\sigma_{\mathcal{V}}^2},
\end{equation}  
where $\mathcal{V}_{SM}$ is the standard theoretical scalar field potential model, $\theta_s$ is the model parameter space, and $\mathcal{V}_{obs}$ are the reconstructed scalar field potential data.

 Moreover, to assess which standard theoretical model provides a better fit to reconstruction $\mathcal{V}(\phi)$, we calculated the reduced values of chi-square ($\chi_{\text{red}}^2$) between the reconstructed $\mathcal{V}(\phi)$ and various standard theoretical models. The $\chi_{\text{red}}$ is computed as:
 \begin{equation}
     \chi^2_{\text{red}} =\frac{1}{N-P} \sum \frac{\left(\mathcal{V}_{SM}(\theta_s,\phi_i) - \mathcal{V}_{obs}(\phi_i)\right)^2}{\sigma_{\mathcal{V}}^2},
 \end{equation}
where $N$ represents the number of data points and $P$ represents the number of parameters in the model. Table \ref{Table 1} presents the reduced chi-square ($\chi^2_{\mathrm{red}}$) values for reconstructed $\mathcal{V}(\phi)$ versus each standard scalar field model, along with their statistical interpretation.

\begin{table*}
\begin{center}
  \caption{Reduced Chi-Square ($\chi^2_{\mathrm{red}}$) Analysis of Reconstructed $\mathcal{V}(\phi)$ Against Standard Scalar Field Models.}
    \label{Table 1}
    \begin{tabular}{l c c }
\hline\hline 
Model      &  $\chi^2_{\mathrm{red}}$ & Interpretation     \\[1ex] \hline\hline
Exponential: $\mathcal{V}_0+\mathcal{V}_1e^{-\lambda\phi}$  & $0.7694$ & Good fit  \\[1ex]
   Power-law: $\mathcal{V}_0+\mathcal{V}_1 \phi^{\lambda}$   & $0.9888$  & Excellent fit  \\[1ex] 
Free Field: $\mathcal{V}_0+ \frac{m^2}{2}\phi^2$ & $1.2150$  & Good fit  \\[1ex] 
Higgs scalar field: $\mathcal{V}_0+\frac{m^2}{2}\phi^2+\frac{\lambda}{4}\phi^4$ & $0.8120$ & Reasonably good fit \\[1ex]
\hline
\end{tabular}
\end{center}
\end{table*}

\subsection{Markov Chain Monte Carlo Method}
We have employed the Markov Chain Monte Carlo (MCMC) technique as a robust Bayesian inference tool to constrain the free parameters of the theoretical model under consideration. Specifically, MCMC is used to determine the best-fit values of the scalar field potential parameters, with the potential reconstruction guided by GP techniques applied to observational data. The MCMC algorithm allows us to efficiently sample the posterior distribution of the model parameters by constructing a Markov chain whose equilibrium distribution corresponds to the posterior probability \( P(\theta | D) \propto \mathcal{L}(D | \theta)\pi(\theta) \), where \( \mathcal{L}(D | \theta) \) is the likelihood of the data \( D \) given the parameters \( \theta \), and \( \pi(\theta) \) denotes the prior distribution. 
The MCMC sampling is carried out using the \texttt{emcee} ensemble sampler, enabling efficient exploration of the parameter space associated with the reconstructed scalar field potentials. This methodology ensures a statistically consistent estimation of the most probable model parameters compatible with current cosmological observations.


To constrain the potential coefficients, we utilize the GP reconstructed $H(z)$ and its associated uncertainties rather than raw observational data, a standard practice in non-parametric reconstruction studies. While this two step approach introduces an intermediate smoothing step that differs from a direct likelihood analysis of the raw data, it provides a practical advantage by enabling stable numerical integration and exploration of the potential's functional form without imposing a specific parametric model a priori. We clarify that our method is non-parametric with respect to the shapes of $H(z)$ and $\mathcal{V}(\phi)$, but not entirely free from theoretical assumptions. Specifically, converting BAO measurements into $H(z)$ requires a fiducial sound horizon scale; we adopt $r_d = 147.09 \pm 0.26$ Mpc, consistent with the Planck 2018 $\Lambda$CDM prediction \citep{P18}, which introduces a mild dependence on early Universe physics and may subtly influence our $H_0$ estimate. Furthermore, the derivation of $\mathcal{V}(\phi)$ explicitly assumes the DBI dynamical equations, meaning our reconstruction is agnostic to the analytical form of the potential rather than fully model-independent. Consequently, the GP method serves as a robust, non parametric tool that yields a well-behaved dataset for stable fitting, and our results should be interpreted as constraints on scalar field potentials within the DBI framework, conditioned on the adopted $r_d$ calibration and GP covariance kernel.




\begin{figure*}[htbp]
\centering

\begin{minipage}{0.45\textwidth}
  \centering
  \includegraphics[width=\linewidth]{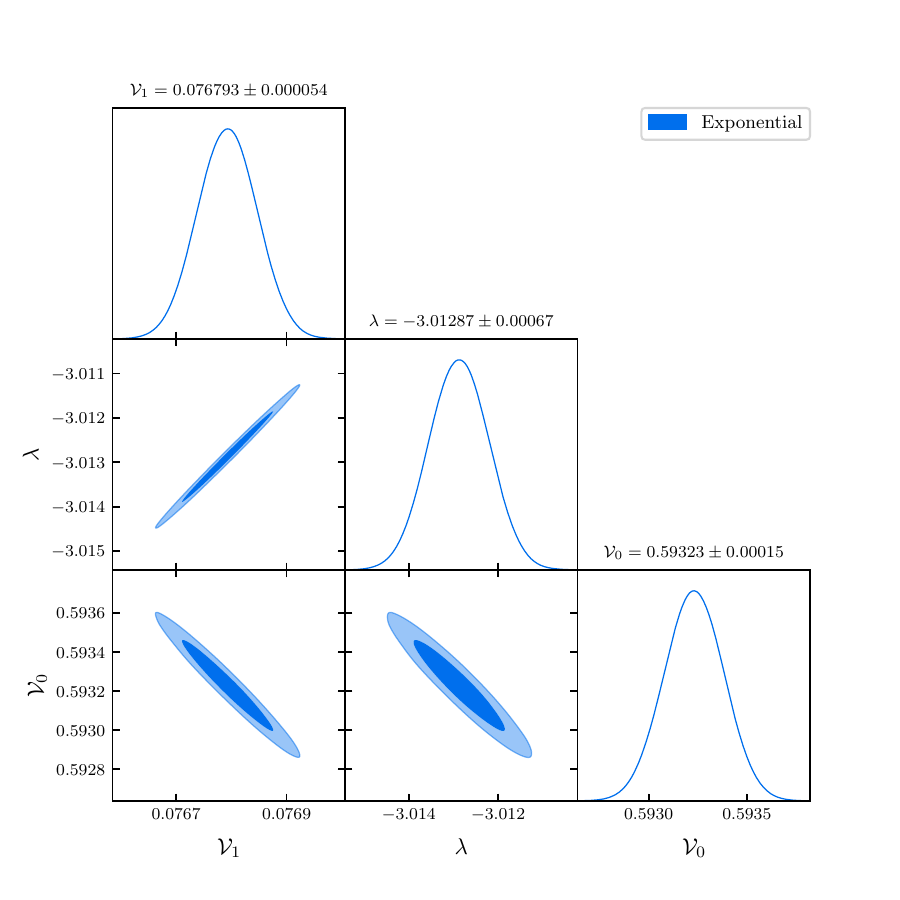}
  \vskip1ex
  \small\textbf{(a)}
\end{minipage}\hspace{0.05\textwidth}%
\begin{minipage}{0.45\textwidth}
  \centering
  \includegraphics[width=\linewidth]{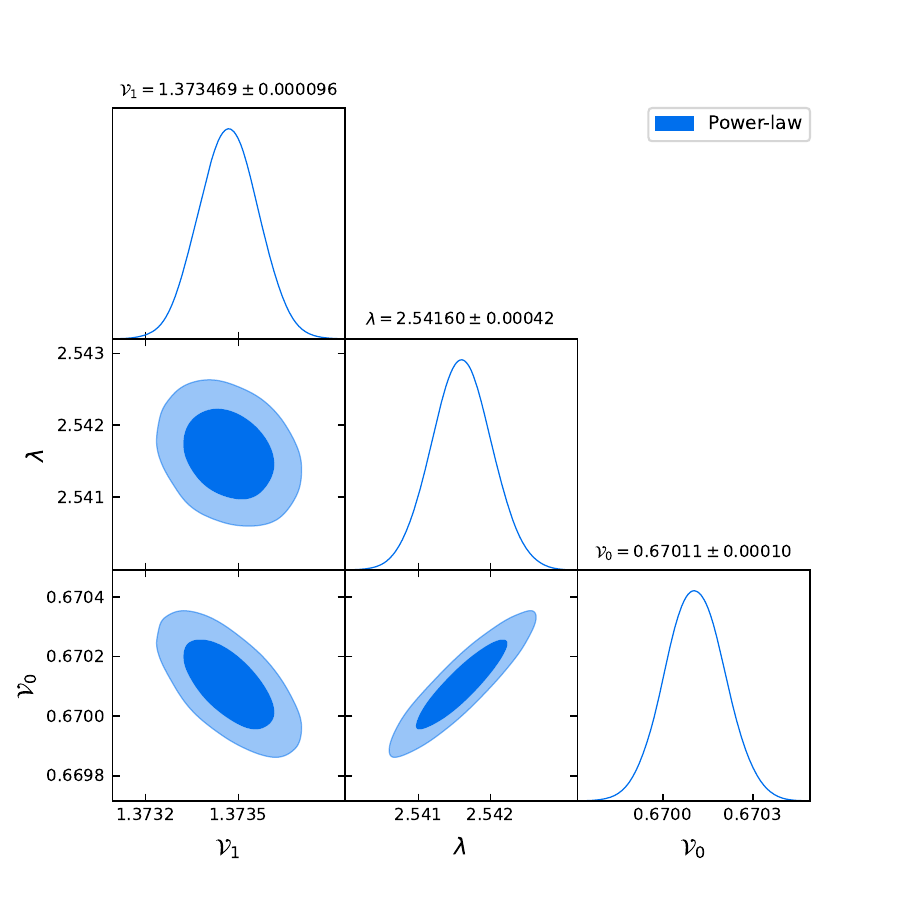}
  \vskip1ex
  \small\textbf{(b)}
\end{minipage}

\vspace{0.4cm}

\begin{minipage}{0.45\textwidth}
  \centering
  \includegraphics[width=\linewidth]{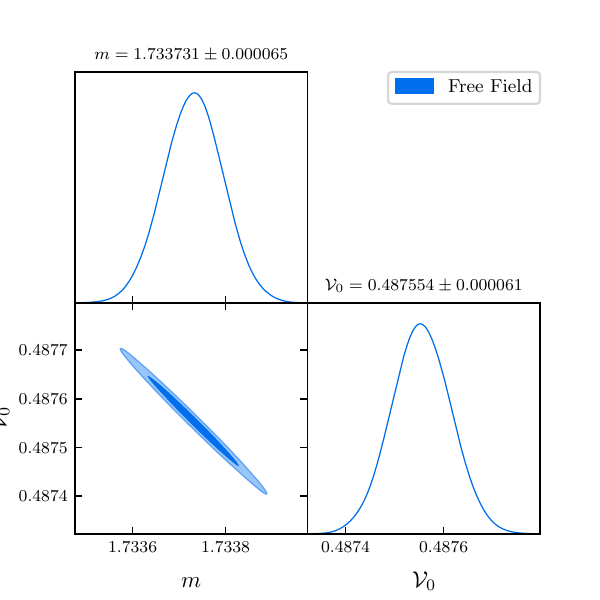}
  \vskip1ex
  \small\textbf{(c)}
\end{minipage}\hspace{0.05\textwidth}%
\begin{minipage}{0.45\textwidth}
  \centering
  \includegraphics[width=\linewidth]{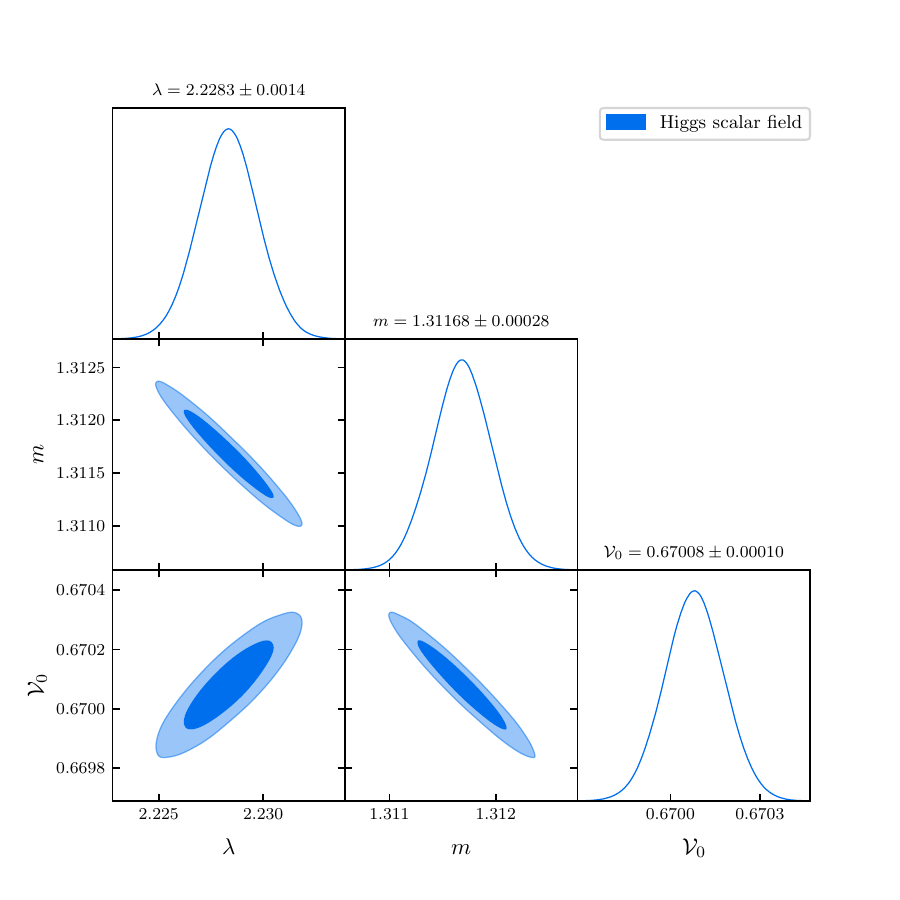}
  \vskip1ex
  \small\textbf{(d)}
\end{minipage}

\caption{\justifying Corner plots showing the correlations and posterior distributions of the free parameters for the Exponential, Power-law, Free Field, and Higgs-like scalar field potentials. The contours represent the $1\sigma$ and $2\sigma$ confidence regions, derived from the reconstructed scalar field potential data.}
\label{fig:05}
\end{figure*}
\subsection{Results and Discussions}
In this work, we consider four different standard scalar field potentials, namely, the Exponential, Power-law, Free field, and Higgs potentials, to compare against our reconstructed dataset. Each potential is fitted to the reconstructed $\mathcal{V}(\phi)$ data shown in Fig.~\ref{fig:04}, and the quality of the fit is assessed using the reduced chi-square statistic. In addition, we perform an MCMC analysis to determine the best-fit values of the model parameters and their corresponding confidence intervals (see Fig.~\ref{fig:05}). In the following, we summarize the properties and interpretations of each model:
\paragraph{\textbf{Exponential potential ($\mathcal{V}_0+\mathcal{V}_1e^{-\lambda\phi}$):}}
 The exponential potential is one of the most studied scalar field potentials, as it is known for its tracker-like behavior, and in recent years it has also provided justification for the swampland conjecture \citep{expnentiallavinia}. In our work, we have reconstructed the potential shown in Fig.~\ref{fig:04} and the exponential potential coefficients to provide a very good estimate of the observational data set. Further, we find that the exponential potential provides a good fit to the reconstructed data, with a reduced chi-square value of $0.7694$. This indicates that the model is statistically consistent with the data and captures its behavior without significant deviations. However, we note that an additional one $\mathcal{V}_0$ is needed for the data to fit. It shows that not just quintessence potential but a nonzero "zero-point" energy is needed for observational consistency and $\mathcal{V}_0$ just mimics the $\Lambda$ term in the $\Lambda$CDM cosmology, while the $\mathcal{V}_1e^{-\lambda \phi}$ denotes the quintessence potential, which is responsible for such a low value of the cosmological constant. It is also evident from the MCMC analysis in Fig.~\ref{fig:05} that there is a negative correlation between $\mathcal{V}_0$ vs $\mathcal{V}_1$ and $\lambda$ that is reasonable for a certain cosmological constant, if $\mathcal{V}_0$ is higher than $\mathcal{V}_1$ would be lower and $\lambda$ would be higher so that overall potential does not change observational $\Lambda$. This is the reason that there is a rather strong correlation between $\mathcal{V}_1$ and $\lambda$, as the sum of the total should be more or less $\Lambda$. However, there is an additional positive correlation between $\mathcal{V}_1$ and $\lambda$ which gives tight constraints on the values of $\lambda$ and $\mathcal{V}_1$.
 \paragraph{\textbf{Power-law potential ($\mathcal{V}_0+\mathcal{V}_1 \phi^{\lambda}$):}} Power-law potential is another widely studied potential in the context of cosmology, as the study of power law potential is very easy in quantum field theory and as Peebles and Ratra have shown \citep{powerlawratra,powerlawratra2}, power-law potential in the de-Sitter background could be a very good candidate for quintessence. As we can see from Fig.~\ref{fig:04}, although this potential, which does not rely on a tracker-like solution, produces an excellent fit to the reconstructed data, with $\chi^2_{\text{red}}=0.9888$. As in the exponential case, the constant term $\mathcal{V}_0$ plays a crucial role in matching the overall energy scale of the potential, again reflecting the need for a nonzero "zero-point" energy. Also, from the MCMC analysis in Fig.~\ref{fig:05} for the power-law potential, as we can see, there is a negative correlation between $\mathcal{V}_1$ and $\mathcal{V}_0$ and $\lambda$, which is again resolvable since the $\Lambda$ for the current universe is fixed from the observation. However, there is a positive correlation between $\mathcal{V}_0$ and $\lambda$ which gives us the constraints on $\lambda$ for a fixed $\mathcal{V}_0$ and $\mathcal{V}_1$. We can see that, in this case, even though there is a correlation between $\mathcal{V}_0$ and $\lambda$, it is not extremely degenerate, as there is much more flexibility in the power potential than in the exponential potential.
\paragraph{\textbf{Free-Field potential ($\mathcal{V}_0+ \frac{m^2}{2}\phi^2$):}} This potential was proposed by Linde \citep{freefieldlinde} as a candidate for chaotic inflation. Later, the free-field study by Urna-Lopez et al. \citep{freefieldlopez} was extended to examine the phenomenology of late-time cosmology. This simple quadratic form also provides a good fit to the reconstructed data as seen in Fig.~\ref{fig:04}, with a reduced chi-square value of $\chi^2_{\text{red}} = 1.2150$. Here, too, the additional $\mathcal{V}_0$ term is needed to reconcile the model with observations, highlighting the role of vacuum energy in driving late-time cosmic acceleration. In this context, we are not concerned with the initial conditions during inflation, as emphasized by Linde \citep{Linde} and others, but rather focus on the universe's late-time dynamics. From the MCMC analysis in Fig.~\ref{fig:05} we see a strong negative correlation between $\mathcal{V}_0$ and $m$, which is consistent with the fact that $\Lambda$ at present is fixed. In this case, we can also see a strong degeneracy as the number of parameters decreases, and it is well known that these free-field models are extremely consistent with the dark energy model due to their tracking-like solutions.
\paragraph{\textbf{Higgs scalar field potential (\(\mathcal{V}_0 + \frac{m^2}{2} \phi^2 + \frac{\lambda}{4} \phi^4\)):}} This potential, which includes both a quadratic and a quartic term, naturally arises from spontaneous symmetry breaking of the $Z_2$ group \citep{higgsamin}. This potential has a very similar feature to the Higgs' mechanism \citep{higgs} in weak interaction, where $U(1)$ gauge symmetry breaking gives the mass of $ W$ and $ Z$  bosons. This potential also has a spontaneous symmetry-breaking mechanism, which can lead to preferred vacuum solutions. With $\chi^2_{\text{red}}=0.8120$, it also fits the reconstructed data reasonably well, and the reconstructed potential is shown in Fig.~\ref{fig:04}. This potential is particularly significant in describing the transition of the universe from one unstable false vacuum to the true vacuum via spontaneous symmetry breaking. As with the previous potentials, an additional \(\mathcal{V}_0\) term is necessary to be consistent with observational data, again indicating the need for a nonzero "zero-point" energy. The \(\mathcal{V}_0\) term mimics the \(\Lambda\) term in the \(\Lambda\)CDM cosmology, while the \(\frac{m^2}{2} \phi^2 + \frac{\lambda}{4} \phi^4\) terms govern the dynamics of the scalar field, describing the transition between vacuum and spontaneous symmetry breaking. For the Higgs-like field, from the MCMC analysis in Fig.~\ref{fig:05} we observe a negative correlation between $m$ and $\mathcal{V}_0$ and $\lambda$, which is expected given that $\Lambda$ is fixed in the current universe. However, here we can also see a positive correlation between $\mathcal{V}_0$ and $\lambda$, which provides a constraint on $\lambda$ for the Higgs field. Just like the power law potential, we can also see here that there is much less degeneracy, unlike exponential and free field models, due to two reasons: first, there are more parameters than free fields, and second, the coefficient of $\phi^4$ that is $\lambda$ is regulating the strong correlations, so in a sense, we can see much less degeneracy than exponential and free field models. Ultimately, these robust, data-driven correlations among the potential parameters are more than just statistical artifacts; they carry direct physical significance. Crucially, they provide essential empirical constraints for constructing high-energy effective field theories (EFTs) that yield DBI scalar fields, ensuring the underlying high-energy dynamics remain consistent with low-energy cosmological observations.
\section{Conclusions}\label{sec6}

In this study, we reconstruct the DE scalar-field potential using the GP method within the Dirac-Born-Infeld (DBI) scalar field framework, employing the combined CC+BAO+DESI dataset. To examine the behavior of DE, we also reconstruct the DE EoS parameter $w(z)$ and the DE density parameter $\Omega(z)$. In this article, our reconstruction is non-parametric with respect to the functional forms of $H(z)$ and $\mathcal{V}(\phi)$, which, for the statistical inference purpose, we have taken the DBI field, though it assumes the DBI dynamical structure and a fiducial $r_d$ calibration for BAO data.


We would also like to note that the $w_{\phi}$ we get from the GP reconstruction is very close to -1, which is not only consistent with the $\Lambda$CDM model but phenomenologically justifies our choice of DBI field, as $w_{\phi}\approx-1$ when the field is very slow rolling.

Using the GP reconstruction, we generated Monte Carlo realizations of $H(z)$ to propagate all reconstruction uncertainties into the inferred Hubble constant. This procedure yields a stable and nearly Gaussian posterior distribution, giving a robust constraint of $H_0 = 69.53 \pm 2.68$ km s$^{-1}$ Mpc$^{-1}$, as shown in Fig.~\ref{fig:01}. The close agreement between the sampled histogram and its best-fitting Gaussian curve indicates that the inference is statistically well behaved. A major advantage of this GP-based approach is its model independence, relying solely on observational data and non-parametric GP priors. The resulting estimate naturally falls between the two leading values involved in the Hubble tension: it is statistically consistent with the Planck $\Lambda$CDM result of $67.4 \pm 0.5$ km s$^{-1}$ Mpc$^{-1}$ at the $\sim0.8\sigma$ level, and likewise consistent with the SH0ES late-universe determination of $73.04 \pm 1.04$ km s$^{-1}$ Mpc$^{-1}$ within $\sim1.3\sigma$. This positions our estimate near the intermediate regime around $H_0 \approx 70$ km s$^{-1}$ Mpc$^{-1}$, a range frequently obtained in model-independent reconstructions and flexible dark-energy scenarios. Hence, the GP method provides a transparent and unbiased determination of $H_0$, delivering a value that is fully acceptable within current observational constraints and offering a balanced perspective that neither exacerbates nor dismisses the existing Hubble tension.


To perform the reconstruction of the scalar field potential, we used reconstructed observational data and a GP prior, along with a choice of covariance function for the GP. 
After running the GP reconstruction, we obtain the reconstructed scalar field potential $\mathcal{V}(\phi)$ in the form of a data set $\mathcal{V}(\phi)$ versus $\phi$, along with its uncertainty $\sigma_{\mathcal{V}}$. To identify the best-fit model for the reconstructed potential, we consider four widely used and physically motivated forms of the potential: the Exponential Potential, the Power-Law Potential, the free field (Quadratic) potential and the Higgs-like Potential. Using the reconstructed $\mathcal{V}(\phi)$ vs. $\phi$ data, we perform a Markov Chain Monte Carlo (MCMC) analysis for each of these potentials. This approach ensures that the fitted parameters for each potential are free from any model-dependent biases that would otherwise arise from assuming a specific cosmological model. Such model-independent reconstructions are particularly relevant in the current cosmological context, where tensions such as the Hubble tension highlight the challenges in identifying the most accurate and predictive models for our universe.

It is worth emphasizing that even though the DBI field was originally proposed as an IR limit for very high energy string theory, it can be used as an alternative to quintessence as it gives observationally consistent cosmological constant \citep{bhagla}, i.e., it recovers $\Lambda$CDM model, as well as it does not hamper the structure formation \citep{paddy2} which can be observationally verified from large scale structure observation or N-Body simulation. Also, as shown by Gorini et al.\citep{gorini}, one can reconstruct the Generalized Chaplygin Gas model from the DBI field, and as Chaplygin gas models are the best phenomenologically fitted model for dark energy models, this shows that the DBI field can be a perfect candidate for dark energy as well.

The four scalar field potentials considered in this study are chosen based on their relevance in current cosmological research and their ability to describe the evolution of DE across cosmic time. These potentials include the \textbf{Exponential Potential} given by \( \mathcal{V}(\phi) = \mathcal{V}_0 + \mathcal{V}_1 e^{-\lambda \phi} \), which provides an excellent fit with \( \chi^2_{\text{red}} = 0.7694 \), suggesting that even dynamical DE models require a baseline cosmological constant-like component for observational consistency. Moreover, this potential aligns well with the Swampland conjecture, indicating that it may be consistent with quantum gravity constraints. The \textbf{Power-law Potential}, \( \mathcal{V}(\phi) = \mathcal{V}_0 + \mathcal{V}_1 \phi^{\lambda} \), yields the best statistical agreement with the reconstruction (\( \chi^2_{\text{red}} = 0.9888 \)). Although it does not rely on a tracker solution, it still captures the key features of cosmic acceleration and DE dynamics, requiring the addition of the \( \mathcal{V}_0 \) term for consistency with the data. The \textbf{Free Field (Quadratic) Potential}, \( \mathcal{V}(\phi) = \mathcal{V}_0 + \frac{1}{2} m^2 \phi^2 \) provides a good fit with \( \chi^2_{\text{red}} = 1.2150 \), although it requires an additional \( \mathcal{V}_0 \) term to match the observational data, highlighting the need for a vacuum energy contribution in late-time cosmology. Lastly, the \textbf{Higgs-Like Potential}, \( \mathcal{V}(\phi) = \mathcal{V}_0 + \frac{1}{2} m^2 \phi^2 + \frac{\lambda}{4} \phi^4 \) fits the data well with \( \chi^2_{\text{red}} = 0.8120 \). This potential is significant in describing transitions between vacua, and, as in the other models, requires the inclusion of \( \mathcal{V}_0 \) to account for the vacuum energy and ensure consistency with observations.
Based on the reduced values of the chi-square ($\chi^2_{\text{red}}$), the potential of the Power-law exhibits the closest agreement with the reconstructed data, indicating an excellent fit. The Exponential and Higgs-like potentials also show good consistency with the reconstruction. Although the Free Field potential yields a slightly higher $\chi^2_{\text{red}}$ value, it still lies within the acceptable range for a good fit. Finally, the analysis shows that all four potential models are compatible with the reconstructed data, with the Power-law potential providing the best match. The inclusion of these four well-motivated potentials enables a comprehensive exploration of how various scalar field dynamics correspond to model-independent reconstructions of the DE potential.

It is worth emphasizing that finding the parameter for the DBI field potential is not just of phenomenological interest, but also carries deep theoretical significance. As mentioned in the introduction, how ``swampland conjectures" \citep{vafa} try to constrain all the cosmological models, especially the scalar field models, by giving a very robust bound on the potential and derivative of the potential. As mentioned, a GP is non-parametric; the potential reconstruction by the GP has been of great interest in the string theory community, as it can be used as a proof to verify the swampland bound for the scalar field potentials as given by the ``string landscape" (particularly dSC or RdSC). So, one can also extend our work regarding the constraints on the DBI field potential and could check the validity of the ``swampland criteria" as done in \citep{elizaldeswampland1,elizaldeswampland2}. Furthermore, the strong correlations observed among the free parameters of the reconstructed potentials (such as $V_1$ and $\lambda$ in the exponential model) have significant physical implications. Because these correlations are derived directly from low-energy cosmological observations, they provide essential empirical constraints for constructing high-energy effective field theories (EFTs) that naturally incorporate DBI scalar fields.
 As far as future work is concerned, one can extend this paper in various directions, such as using a more general K-essence field beyond the DBI field to constrain the potentials using GP. One can also verify the swampland-like conjectures in DBI fields and other K-essence fields to comment on the consistency of the observation and string theory prediction. Last but not least, one can also extend this study to general teleparallel gravity theories, such as  $f(T)$ gravity with the torsion scalar $T$ and $f(Q)$ gravity with the non-metricity scalar $Q$, in order to check the consistency of such modified gravity frameworks with current observational data.\\

\textbf{Data availability:} No new data are associated with this article.

\section*{Acknowledgments}
 SG acknowledges the Council of Scientific and Industrial Research (CSIR), Government of India, New Delhi, for the Junior Research Fellowship (File no.09/1026(13105)/2022-EMR-I).
Gaurav N. Gadbail and Kazuharu Bamba acknowledge the support by the JSPS KAKENHI Grant Number 25KF0176. The work of Kazuharu Bamba was supported in part by the JSPS KAKENHI Grant Number 24KF0100 and Competitive Research Funds for Fukushima University Faculty (25RK011). PKS thanks IUCAA, Pune, India, for its assistance through the visiting associateship program and the Anusandhan National Research Foundation, Department of Science and Technology, Government of India, for financial support to carry out Research project No.: CRG/2022/001847.

\end{document}